\documentclass[12pt]{article}
\usepackage{a4wide}
\usepackage{amsmath,amssymb,bbm}
\usepackage{multirow}

\def\a{\alpha}
\def\b{\beta}
\def\g{\gamma}

\def\d{\delta}
\def\e{\epsilon}
\def\m{\mu}
\def\n{\nu}
\def\r{\rho}
\def\s{\sigma}
\def\de{\partial}

\begin{document}
\setcounter{page}{1}
%


%

\def\pct#1{(see Fig. #1.)}

\begin{titlepage}
\hbox{\hskip 12cm KCL-MTH-10-02  \hfil}
\vskip 1.4cm
\begin{center}  {\Large  \bf  The very-extended trombone}

\vspace{1.8cm}

{\large \large Fabio Riccioni } \vspace{0.8cm}

{\sl Department of Mathematics\\
\vspace{0.3cm}
 King's College London  \\
\vspace{0.3cm}
Strand \ \  London \ \ WC2R 2LS \\
\vspace{0.3cm} UK} \\
\vspace{.8cm} {E-mail: {\tt Fabio.Riccioni@kcl.ac.uk}}
\end{center}
\vskip 1.5cm

\abstract{Starting from the very-extended Kac-Moody algebra
$E_{11}$, we consider the algebra $E_{11,D}^{local}$, obtained by
adding to the non-negative level $E_{11}$ generators the
$D$-dimensional momentum operator and an infinite set of additional
generators that promote the global $E_{11}$ symmetries to gauge
ones. We determine all the possible trombone deformations of this
algebra, that is the deformations that involve the $D$-dimensional
scaling operator. The Jacobi identities imply that such deformations
are uniquely determined by a single tensor belonging to the same
representation of the internal symmetry group as the vector
generators and satisfying additional quadratic constraints. The
non-linear realisation of the deformed algebra gives the field
strengths of the theory which are those of any possible maximal
supergravity theory in which the global scaling symmetry is gauged
in any dimension. All the possible deformed algebras are in one to
one correspondence with all such maximal supergravity theories. The
tensor that parametrises the deformation is identified with the
embedding tensor that is used to parametrise all maximal
supergravity theories with gauged scaling symmetry, and the
quadratic constraints that we determine exactly coincide with the
field theory results. }

\vfill
\end{titlepage}
\makeatletter \@addtoreset{equation}{section} \makeatother
\renewcommand{\theequation}{\thesection.\arabic{equation}}
\addtolength{\baselineskip}{0.3\baselineskip}

\section{Introduction}
The hidden symmetries of ungauged maximal supergravities have played
a crucial role in our understanding of string theory dualities.
These theories in dimension lower than ten are unique and can be
obtained by torus dimensional reduction from both eleven dimensional
\cite{sugra11} and IIB \cite{IIB} supergravities. For instance, in
four dimensions the theory develops an $E_{7(7)}$ symmetry
\cite{cremmerjuliaD=4}, while the hidden symmetry of the
five-dimensional theory is $E_{6(6)}$ \cite{sugraD=5}. In general we
refer to the internal symmetry group of the ungauged maximal
$D$-dimensional supergravity theory as the Cremmer-Julia group, and
we denote it by $E_{11-D(11-D)}$.

Gauged maximal supergravities generally arise as deformations of the
ungauged ones by imposing supersymmetry together with gauge
invariance with respect to a subgroup of $E_{11-D(11-D)}$. For
instance the four dimensional ${\cal N}=8$ theory of
\cite{dewitnicolai} is a deformation of the massless maximal
supergravity of \cite{cremmerjuliaD=4} where an $SO(8)$ subgroup of
$E_{7(7)}$ is gauged. Relatively recently, all maximal gauged
supergravity theories in each dimension $D$ have been classified in
terms of a single object called the embedding tensor which can be
thought of as belonging to a representation of the internal symmetry
group $E_{11-D(11-D)}$ of the supergravity theory in $D$ dimensions
\cite{nicolaisamtlebenD=3,dWSTgeneral,dWSTD=5,samtlebenweidnerD=7,dWSTmagnetic,dWSTD=4,embeddingtensorD=6}.
Supersymmetry specifies the $E_{11-D}$ representation of the
embedding tensor by imposing a set of linear (or representation)
constraints, while imposing that the embedding tensor is a constant
 breaks the $E_{11-D}$ symmetry to the subgroup the embedding tensor is an
invariant tensor of. This subgroup is indeed the gauge group, and
consistency imposes additional quadratic constraints on the
embedding tensor, which can be viewed as the Jacobi identities for
the structure constants of this gauge group.

Maximal supergravity theories have a very elegant classification in
terms of the very-extended infinite-dimensional Kac-Moody algebra
$E_{11}$ \cite{E11}. This algebra was first conjectured in
\cite{E11} to be a symmetry of M-theory. The maximal supergravity
theory in $D$ dimensions corresponds to decomposing $E_{11}$ in
terms of $GL(D,\mathbb{R}) \otimes E_{11-D}$, and thus the
occurrence of the internal symmetry $E_{11-D}$ appears natural from
this perspective. The $GL(D,\mathbb{R})$ group corresponds to the
constant part of the diffeomorphism group in $D$ dimensions, and
gravity is described in this framework as the non-linear realisation
of the diffeomorphism group with the $D$ dimensional Lorentz group
as local subgroup
\cite{ogievetsky,borisovogievetsky,petersuperconformal}. For
instance, the IIA theory naturally has from the $E_{11}$ viewpoint
an $\mathbb{R}^+$ symmetry corresponding to the shift of the
dilaton. Decomposing the adjoint representation of $E_{11}$ with
respect to the subalgebra associated to the IIA theory one obtains
generators that are associated to the IIA fields and their duals
\cite{E11}. One also finds a generator with nine antisymmetric
ten-dimensional spacetime indices, which is associated to a 9-form
\cite{axeligorpeter}. This 9-form has a 10-form field strength,
which can be thought as the dual of the mass parameter of Romans.
Therefore the Romans massive IIA is naturally encoded in $E_{11}$
\cite{igorpeterromans}.

More generally, decomposing the $E_{11}$ algebra in a given
dimension and considering only the level zero generators (that is
the generators of $GL(D,\mathbb{R}) \otimes E_{11-D}$, that are
associated to the graviton and the scalars) and the positive level
generators with completely antisymmetric indices, that are
associated to forms, one finds in all cases the field content of the
$D$-dimensional supergravity theory, in a democratic formulation in
which all fields appear together with their magnetic duals
\cite{axeligorpeter,fabiopeterE11origin}. One also finds $D-1$ form
generators in representations of $E_{11-D}$ which remarkably are the
same as those fixed by the linear constraints on the embedding
tensor for the maximal gauged supergravities in dimension $D$
\cite{fabiopeterE11origin,ericembeddingtensor}. Exactly like in the
case on Romans, one thinks of the $D-1$ form fields as being dual to
the embedding tensor, obtaining in this way a classification of all
possible maximal gauged supergravities in terms of $E_{11}$.

The action of positive level $E_{11}$ generators with completely
antisymmetric spacetime indices in the non-linear realisation
corresponds to gauge transformations for the associated form fields
that are linear in the spacetime coordinates. In
\cite{fabiopeterogievetsky} this algebra was enlarged in order to
include arbitrary gauge transformations. The resulting algebra
includes the non-negative level generators as well as momentum and
an infinite set of additional generators, that were called
Ogievetsky, or Og generators, that correspond to an expansion in the
spacetime coordinates of the gauge parameters. This extension is
dimension-dependent, and it was called $E_{11,D}^{local}$ in
\cite{fabiopeterogievetsky}. From the non-linear realisation of the
$E_{11,D}^{local}$ algebra with as local subalgebra the $D$
dimensional Lorentz algebra times the maximal compact subalgebra of
$E_{11-D}$ one computes all the field strengths of the massless
maximal supergravity in $D$ dimensions.

Given the local $E_{11}$ algebra in $D$ dimensions, one can consider
its massive deformations. In \cite{hierarchyE11} the deformations
that do not involve the $GL(D, \mathbb{R})$ generators were studied,
and the consistency of the deformed algebra implies that all
possible deformations are parametrised by a constant quantity that
turns out to be the embedding tensor. Remarkably, the Jacobi
identities impose constraints on the embedding tensor that are
exactly the linear and quadratic constraints that were obtained in
the field theory analysis. All the possible deformations are thus in
one to one correspondence with all the possible gauged
supergravities resulting from the gauging of a subgroup of
$E_{11-D}$, while the Maurer-Cartan form gives all the field
strengths in a straightforward way.

There are maximal gauged supergravities that do not arise from the
gauging of a subgroup of the internal symmetry, but correspond to
the gauging of the global scaling symmetry, also called ``trombone''
symmetry, that leaves the field equations invariant, but rescales
the action. The fact that the scaling symmetry is not a symmetry of
the lagrangian implies that the corresponding gauged theory does not
admit a lagrangian formulation, but only field equations. The first
example of such a theory is the gauged IIA theory of
\cite{hlw,fibrebundles}, while the nine-dimensional analogue was
discussed in \cite{allgaugingsD=9}. Recently, in \cite{henning} a
systematic classification of these theories in dimension from three
to six was provided. This was achieved by the introduction of a new
type of embedding tensor, belonging to the $E_{11-D}$ representation
which is conjugate to the one of the vector fields. The consistency
of the gauge algebra implies additional quadratic constraints for
this embedding tensor.

In \cite{trombone1} it was shown that the local $E_{11}$ algebra
admits a new type of deformations that are associated to maximal
supergravities in which the trombone symmetry is gauged. These
deformations involve the generator of scale transformations, which
is the trace of the $GL(D,\mathbb{R})$ generators. This deformation
was discussed in detail for the IIA case, corresponding to the
gauged IIA theory of \cite{hlw,fibrebundles}.

In this paper we perform a systematic analysis of these deformations
in any dimension from three to nine. In particular, we show that the
deformations are parametrised by a constant quantity in the same
$E_{11-D}$ representation as the vector generator, which is
identified with the embedding tensor of \cite{henning}. We show that
the Jacobi identities imply quadratic constraints for the embedding
tensor that are exactly equivalent to those derived in
\cite{henning} for the cases from three to six dimensions, and we
also derive the quadratic constraints in seven, eight and nine
dimensions. Moreover, from the non-linear realisation we derive the
field strengths and gauge transformations of the fields.

It is important to stress that the local $E_{11}$ algebra is not
compatible with the full $E_{11}$ symmetry, and only its $GL(D
,\mathbb{R}) \otimes E_{11-D}$ subalgebra survives the introduction
of momentum and the Og generators in $D$ dimensions. A similar
argument applies to all the deformed cases, where $E_{11-D}$ is
further broken by the embedding tensor to the gauge subgroup.
$E_{11}$ is considered throughout this paper as the universal
algebraic framework to describe the gauge algebra (including
gravity) of all maximal supergravity theories.

The paper is organised as follows. In section 2 we derive the
general method of constructing the local $E_{11}$ algebra associated
to the trombone deformations in any dimension. In section $D$, with
$D=3, ..., 9$, we explicitly derive the deformed algebra in a given
dimension $D$. Section 10 contains the conclusions.

\section{The general method}
In this section we perform a general analysis which will then be
used from sections 3 to 9 in each dimension separately. The analysis
is based on the results of \cite{trombone1}, where it was shown how
one can deform the local $E_{11}$ algebra corresponding to IIA
supergravity to obtain a non-linear realisation corresponding to the
gauged IIA theory of \cite{hlw,fibrebundles}.

In \cite{hierarchyE11} it was shown that all gauged supergravities
that result from the gauging of a subgroup of the internal symmetry
group arise from all possible deformations of the local $E_{11}$
algebra in $D$ dimensions that do not involve the $GL(D,
\mathbb{R})$ generators. In order to show this, in section 2 of
\cite{hierarchyE11} a general notation was introduced to discuss in
a single framework all different dimensions. In this paper we
consider in any dimension deformations that also involve the $GL(D,
\mathbb{R})$ generators, and in this section we develop a general
formalism which is the analogue for these deformations to section 2
of \cite{hierarchyE11}. In particular all the notations are taken
from there. We thus decompose $E_{11}$ in terms of $GL(D,
\mathbb{R}) \otimes E_{11-D}$. This corresponds to deleting node $D$
in the $E_{11}$ Dynkin diagram of fig. 1.
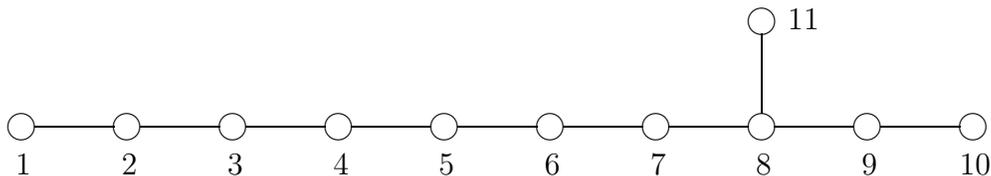
\begin{figure}[h]
\begin{center}
\begin{picture}(380,70)
\multiput(10,10)(40,0){6}{\circle{10}}
\multiput(250,10)(40,0){3}{\circle{10}} \put(370,10){\circle{10}}
\multiput(15,10)(40,0){9}{\line(1,0){30}} \put(290,50){\circle{10}}
\put(290,15){\line(0,1){30}} \put(8,-8){$1$} \put(48,-8){$2$}
\put(88,-8){$3$} \put(128,-8){$4$} \put(168,-8){$5$}
\put(208,-8){$6$} \put(248,-8){$7$} \put(288,-8){$8$}
\put(328,-8){$9$} \put(365,-8){$10$} \put(300,47){$11$}
\end{picture}
\caption{\sl The $E_{11}$ Dynkin diagram. }
\end{center}
\end{figure}
In the decomposition of the adjoint representation of $E_{11}$ in
terms of representations of $GL(D, \mathbb{R}) \otimes E_{11-D}$, we
are only interested in the level zero generators, that is the
generators of $GL(D, \mathbb{R})$ and those of $E_{11-D}$, and the
positive level generators with completely antisymmetric  $GL(D,
\mathbb{R})$ indices. These are
  \begin{equation}
  K^\m{}_\n \quad R^\alpha \quad R^{\m_1 , M_1} \quad R^{\m_1 \m_2 ,
  M_2} \quad ... \quad .
  \end{equation}
The generator $R^{\m_1 ...\m_n , M_n}$ carries the representation
${\bf R_n}$ of $E_{11-D}$ which transforms the $M_n$ index. The
$E_{11-D}$ generators in such representation are
$D^\alpha_{N_m}{}^{M_m}$. The $E_{11}$ algebra involving the form
generators is
  \begin{equation}
  [R^{\m_1\ldots \m_m, M_m},R^{\n_1\ldots \n_n,
  N_n}]=f^{M_m N_n}{}_{P_{m+n}} R^{\m_1\ldots \m_{m}\n_1 \ldots \n_n,{P_{m+n}}
  } \label{undeformedE11}
  \end{equation}
and
  \begin{equation}
  [ R^\alpha, R^{\m_1\ldots \m_m, M_m} ]= (D^\alpha)_{N_m}{}^{M_m}
  R^{\m_1\ldots \m_m, N_m} \quad ,\qquad
  [R^\alpha,R^\beta]=f^{\alpha\beta}{}_\gamma R^\gamma \quad  ,
  \label{undeformedscalar}
  \end{equation}
where $f^{\a\b}{}_\g$ are the structure constants of $E_{11-D}$ and
$f^{M_m N_n}{}_{P_{n+m}}$ are generalised structure constants.

While the scalar generators are associated to the global symmetry
$E_{11-D}$, all the form generators are associated to gauge fields,
and as explained in reference \cite{fabiopeterogievetsky}, the
global symmetry associated to the form generators is promoted to a
local one by the addition of the spacetime translation operator
$P_\m$ and an infinite number of so called Ogievetsky, or Og,
generators. In fact for our purposes we need only add the lowest
order, or Og 1, such generators, $K^{\m ,\n_1\ldots \n_n, M_n}$,
which by definition obey the commutator
  \begin{equation}
  [ K^{\m ,\n_1\ldots \n_n, M_n},P_\r ]= \delta_\r^{\m}R^{\n_1\ldots \n_n, M_n}
  -\delta_\r^{[\m}R^{\n_1\ldots \n_n ], M_n}\quad .
  \label{undeformedKwithP}
  \end{equation}
Each generator $K^{\m,\n_1\ldots \n_n, M_n}$ is associated with the
$E_{11}$ generator $R^{\m_1\ldots \m_n, M_n}$ and carries the same
internal symmetry representation, ${\bf R_n}$. It also satisfies
$K^{ [\m,\n_1\ldots \n_n ], M_n }=0$.

The deformations of the above $E_{11,D}^{local}$ algebra considered
in \cite{hierarchyE11} are associated to gaugings of subgroups of
the internal symmetry group $E_{11-D}$. As such, they do not involve
the $GL(D, \mathbb{R})$ generators, which commute with the
$E_{11-D}$ generators. Correspondingly, these deformations leave the
$D$ dimensional metric invariant. Here we want to consider a new
class of deformations, and we take as our starting point the
deformed commutator
  \begin{equation}
  [ R^{\m ,M_1} , P_\n ]  =  \delta^\m_\n ( \Theta^{M_1} K + a
  \Theta^{N_1} g_{\a\b} D^\a_{N_1}{}^{M_1} R^\beta ) \quad ,
  \label{defalgebra1}
  \end{equation}
where the parameter $a$ is to be determined, and we denote with $K$
the trace of the $GL(D, \mathbb{R})$ generators $K^\m{}_\n$,
  \begin{equation}
  K = K^\m{}_\m \quad ,
  \end{equation}
whose commutator with the form generators is
  \begin{equation}
  [ K , R^{\m_1 ...\m_m , M_m} ] = m R^{\m_1 ...\m_m , M_m} \quad .
  \end{equation}
It can be shown that an additional term of the form
  \begin{equation}
  [ R^{\m ,M_1} , P_\n ] = \Theta^{M_1} K^\m{}_\n \quad ,
  \end{equation}
which involves the $GL(D, \mathbb{R})$ generator and not just its
trace, can be reabsorbed in a redefinition of $R^{\m ,M_1}$ of the
form
  \begin{equation}
  R^{\m ,M_1} \rightarrow R^{\m ,M_1} -2 \Theta^{M_1} K^{\m\n}{}_\n
  \quad ,
  \end{equation}
where $K^{\m\n}{}_\r$ is the Og 1 gravity generator. Therefore, we
will not consider this term in this paper.

These deformations are therefore characterised by an embedding
tensor in the same representation ${\bf R_1}$ as the 1-form
generators. The consistency of the algebra will then in general
imply that the commutator of the $n$-form generator with momentum
does not vanish, and we thus write
  \begin{equation}
  [ R^{\m_1 ...\m_n , M_n }, P_\n ] =  \Theta^{M_1} S^{M_n}{}_{M_1
  M_{n-1}} \delta^{[\m_1}_\n R^{\m_2 ...\m_n ], M_{n-1}}
  \label{defalgebra2}
  \end{equation}
where $S^{M_n}{}_{M_1 M_{n-1}}$ is an invariant tensor of
$E_{11-D}$. This defines the trombone deformation of the  algebra
${E}_{11,D}^{local}$. The closure of the algebra also implies that
the commutators of eqs. (\ref{undeformedE11}) and
(\ref{undeformedKwithP}) become
  \begin{eqnarray}
  & & [R^{\m_1\ldots \m_m, M_m},R^{\n_1\ldots \n_n,
  N_n}]=f^{M_mN_n}{}_{P_{m+n}} R^{\m_1\ldots \m_{m}\n_1 \ldots \n_n,{P_{m+n}}
  } \nonumber \\
  & & \quad \quad + \Theta^{M_1} T^{M_m N_n}{}_{M_1 M_{m+n-1}} K^{[\m_1
  , \m_2 ... \m_n ] \n_1 ...\n_m , M_{m+n -1} } \quad , \nonumber \\
  & & [ K^{\m , \n_1 ...\n_m , M_m } , P_\r ] = \delta^\m_\r R^{\n_1
  ...\n_m , M_m } - \delta^{[ \m}_\r R^{\n_1
  ...\n_m ], M_m } \nonumber \\
  & & \quad \quad +  \Theta^{N_1} U^{M_m}{}_{N_1 P_{m-1} }
  \delta^{[\n_1}_\r K^{| \m ,| \n_2 ... \n_m ] , P_{m-1}} \quad .
  \label{defalgebra3}
  \end{eqnarray}
where $T^{M_m N_n}{}_{M_1 M_{m+n-1}}$ and $U^{M_m}{}_{N_1 P_{m-1} }$
are invariant tensors of $E_{11-D}$.

The tensor  $\Theta^{M_1}$  is clearly not an invariant tensor of
$E_{11-D}$, given that it transforms in the ${\bf R_1}$
representation. Therefore assuming that the embedding tensor
$\Theta^{M_1}$ is constant breaks the $E_{11-D}$ symmetry. For this
to be consistent,  $\Theta^{M_1}$ must satisfy quadratic
constraints. Starting from the algebra of eqs. (\ref{defalgebra1}),
(\ref{defalgebra2}) and (\ref{defalgebra3}), as well as eq.
(\ref{undeformedscalar}), and imposing the closure of the Jacobi
identities, we now determine the conditions that the invariant
tensors $S^{M_n}{}_{M_1 M_{n-1}}$, $T^{M_m N_n}{}_{M_1 M_{m+n-1}}$
and $U^{M_m}{}_{N_1 P_{m-1} }$ must satisfy, as well as the
quadratic constraints on the embedding tensor $\Theta^{M_1}$. As we
will see in sections from 3 to 6, these constraints are exactly the
constraints determined in \cite{henning} from a field theory
analysis in dimensions from 3 to 6. In sections 7, 8 and 9 we will
also determine explicitly the constraints in dimensions 7, 8 and 9.
The ten-dimensional case, corresponding to the gauged IIA theory of
\cite{hlw,fibrebundles}, was analysed in \cite{trombone1}, where it
was also shown how the ten-dimensional deformed algebra arises from
an eleven-dimensional perspective.

We first consider the Jacobi identity involving the operators
$R^{\m, M_1}$, $R^{\n, N_1}$ and $P_\r$. We get two conditions,
coming from the antisymmetric and symmetric terms in $\m \n$
respectively, which are
  \begin{eqnarray}
  & & f^{M_1 N_1}{}_{P_2} S^{P_2}{}_{Q_1 P_1} = 2 [
  \delta^{( M_1}_{Q_1} \delta^{N_1 )}_{P_1} + a D_{\alpha Q_1}{}^{(
  M_1} D^{\alpha}_{P_1}{}^{N_1 )} ]\nonumber \\
  & &  T^{M_1 N_1 }{}_{Q_1 P_1} = 2 [\delta^{[ M_1}_{Q_1} \delta^{N_1 ]}_{P_1} + a D_{\alpha
  Q_1}{}^{[
  M_1} D^{\alpha}_{P_1}{}^{N_1 ]} ] \quad .
  \label{SandTareinvarianttensors}
  \end{eqnarray}
These two conditions determine $S^{P_2}{}_{Q_1 P_1}$ and $T^{M_1 N_1
}{}_{Q_1 P_1}$, and as we will see in all cases in the next
sections, they also determine independently the parameter $a$.

Proceeding this way one determines all the conditions that the
invariant tensors in eqs. (\ref{defalgebra2}) and
(\ref{defalgebra3}) satisfy. Given that our goal will be the
computation of the field strengths and gauge transformations of the
fields, we are only interested in the part of the algebra that
involves the deformed $E_{11}$ generators. Therefore we will from
now on neglect the contribution to the Jacobi identities coming from
the Og generators, and always assume that the upstairs spacetime
indices are completely antisymmetrised. This is exactly the same
attitude that was taken in \cite{hierarchyE11}  when computing the
field strengths and gauge transformations for the deformed algebras
associated to  gauged supergravities in which an internal symmetry
is gauged. Considering only the antisymmetric contribution, the
Jacobi identity involving $R^{\m ,M_1}$, $R^{\n_1 ...\n_n , N_n}$
and $P_\r$ gives
  \begin{equation}
  n \delta_{P_1}^{M_1} \delta^{N_n}_{R_n} + a D_{\alpha ,
  P_1}{}^{M_1} D^\alpha_{R_n}{}^{N_n} = S^{N_n}{}_{P_1 Q_{n-1}}
  f^{M_1 Q_{n-1}}{}_{R_n} + f^{M_1 N_n}{}_{S_{n_1}}
  S^{S_{n+1}}{}_{P_1 R_n} \quad .
  \end{equation}
Starting from the first of eqs. (\ref{SandTareinvarianttensors}),
which determines $S^{P_2}{}_{Q_1 P_1}$, this equation determines by
induction all the invariant tensors $S^{N_n}{}_{P_1 Q_{n-1}}$ in eq.
(\ref{defalgebra2}). This is all we need to compute the field
strengths and gauge transformations of the fields. Nonetheless, it
is important to stress that the consistency of the algebra imposes
in general the presence of the Og generators in the first of eqs.
(\ref{defalgebra3}), and the invariant tensors  $T^{M_m N_n}{}_{M_1
M_{m+n-1}}$ and $U^{M_m}{}_{N_1 P_{m-1} }$ can easily be determined
requiring the closure of the Jacobi identities and considering the
terms which are not completely antisymmetric in their spacetime
indices.

We then consider the quadratic constraints. These arise from Jacobi
identities involving a deformed $E_{11}$ generator and two momenta,
or a deformed $E_{11}$ generator, the momentum operator and the
scalar operator $\Theta^{M_1} K + a \Theta^{N_1} D_{\alpha ,
N_1}{}^{M_1} R^\alpha$. The Jacobi identity involving the vector
generator $R^{\m , M_1}$ and two momenta imposes that
  \begin{equation}
  [ \Theta^{M_1} K + a \Theta^{N_1} D_{\a , N_1}{}^{M_1} R^\a , P_\m ] = 0 \quad . \label{KwithP}
  \end{equation}
This implies that the deformed commutation relation
  \begin{equation}
  [ \Theta^{N_1} D_{\a , N_1}{}^{M_1} R^\a , P_\m ] = \frac{1}{a}
  \Theta^{M_1} P_\m
  \end{equation}
holds. The Jacobi identity involving $R^{\m_1 \m_2, M_2}$ and two
momentum operators gives
  \begin{equation}
  \Theta^{M_1 } \Theta^{N_1} S^{M_2}{}_{M_1 N_1} K + a \Theta^{M_1}
  \Theta^{P_1} S^{M_2}{}_{M_1 N_1} D_{\alpha , P_1}{}^{N_1} R^\alpha
  = 0 \quad .
  \end{equation}
The term proportional to $K$ vanishes provided that $\Theta^{M_1}$
satisfies the constraint
  \begin{equation}
  \Theta^{M_1 } \Theta^{N_1} S^{M_2}{}_{M_1 N_1} =0\quad ,
  \label{purespinorconstraint}
  \end{equation}
which is precisely the generalised ``pure spinor constraint'' found
in \cite{henning} in $D=3,4,5,6$. Using the invariance of
$S^{M_2}{}_{M_1 N_1}$, this constraint means that the product of two
embedding tensors vanishes when projected on the representation
${\bf R_2}$ to which the 2-form generators belong. The term
proportional to $R^\alpha$ also vanishes using eq.
(\ref{purespinorconstraint}) and the invariance of $ S^{M_2}{}_{M_1
N_1}$. The Jacobi identity involving $R^{\m_1 ...\m_n , M_n}$, with
$n>1$, and two momentum operators gives
  \begin{equation}
  \Theta^{N_1} \Theta^{P_1} S^{M_n}{}_{N_1 Q_{n-1} }
  S^{Q_{n-1}}{}_{P_1 R_{n-2}} =0 \quad .
  \end{equation}
As will be clear in the next sections, this constraint derives in
all cases from eq. (\ref{purespinorconstraint}) and the condition of
$E_{11-D}$ invariance of $S^{M_n}{}_{N_1 Q_{n-1} }$.

We now consider the quadratic constraints resulting from the Jacobi
identity involving a deformed $E_{11}$ generator, the momentum
operator and the scalar operator $\Theta^{M_1} K + a \Theta^{N_1}
D_{\alpha , N_1}{}^{M_1} R^\alpha$. We first consider the case in
which the $E_{11}$ generator is $R^{\m ,M_1}$, and we get the
equation
  \begin{eqnarray}
  & & [ \Theta^{M_1} \Theta^{P_1} +a \Theta^{N_1} \Theta^{Q_1}
  D_{\alpha, N_1}{}^{M_1} D^\alpha_{Q_1}{}^{P_1} ] K \nonumber \\
  & & + a [
  \Theta^{M_1} \Theta^{Q_1} D_{\alpha , Q_1}{}^{P_1} + a
  \Theta^{N_1} \Theta^{S_1} D_{\beta , N_1}{}^{M_1}
  D^\beta_{S_1}{}^{Q_1} D_{\alpha , Q_1}{}^{P_1} ] R^\alpha =0 \quad
  .
  \end{eqnarray}
Both the $K$ and the $R^\beta$ term give the condition
  \begin{equation}
  \Theta^{N_1} \Theta^{Q_1} [ \delta^{M_1}_{N_1} \delta^{P_1}_{Q_1}
  + a D_{\alpha, N_1}{}^{M_1} D^\alpha_{Q_1}{}^{P_1} ] = 0 \quad .
  \label{parametera}
  \end{equation}
As we have already mentioned, the parameter $a$ is determined by
imposing the closure of the Jacobi identity involving two 1-form
generators and the momentum operator. We will show in the next
sections that using this value of $a$, eq. (\ref{parametera}) is
automatically implied by eq. (\ref{purespinorconstraint}) in all
dimensions.

The last set of quadratic constraints come from the Jacobi identity
involving $R^{\m_1 ...\m_m, M_m}$, momentum and the scalar operator.
This gives
  \begin{eqnarray}
  & & \Theta^{N_1} \Theta^{Q_1} [ \delta^{M_1}_{N_1} S^{P_n}{}_{Q_1
  R_{n-1} } -a S^{P_n}{}_{Q_1 Q_{n-1}} D_{\alpha , N_1}{}^{M_1}
  D^\alpha_{R_{n-1}}{}^{Q_{n-1}} \nonumber \\
  & & +a S^{Q_n}{}_{Q_1 R_{n-1}}
  D_{\alpha , N_1 }{}^{M_1} D^\alpha_{Q_n}{}^{P_n}] = 0 \quad ,
  \end{eqnarray}
and using the condition of invariance of $S^{P_n}{}_{Q_1
  R_{n-1} }$ this becomes
    \begin{equation}
    \Theta^{N_1} \Theta^{Q_1} [ \delta^{M_1}_{N_1} S^{P_n}{}_{Q_1
  R_{n-1} } + a S^{P_n}{}_{P_1 R_{n-1}} D_{\alpha , N_1}{}^{M_1}
  D^\alpha_{Q_1}{}^{P_1} ] = 0 \quad ,
  \end{equation}
which is eq. (\ref{parametera}) contracted by $S^{P_n}{}_{P_1
R_{n-1}}$. We have thus determined all quadratic constraints, and we
will show in the following sections that they are all automatically
satisfied provided that eq. (\ref{purespinorconstraint}) holds and
that $a$ is fixed accordingly.

In \cite{henning} the trombone deformations were considered together
with the deformations resulting from the gauging of a subgroup of
the internal symmetry group, and quadratic constraints involving the
embedding tensors associated to both deformations were derived in
all dimensions from 3 to 6. If the quadratic constraints allow for
the simultaneous presence of both embedding tensors, this would
imply the presence of a new class of theories with simultaneously
non-vanishing embedding tensors. This was indeed the outcome of the
analysis in \cite{henning} in any dimension. From our algebraic
viewpoint, this corresponds to considering as starting point the
more general commutator
  \begin{equation}
  [ R^{\m ,M_1} , P_\n ]  =  \delta^\m_\n ( \Theta^{M_1} K + a
  \Theta^{N_1} g_{\a\b} D^\a_{N_1}{}^{M_1} R^\beta + \Theta^{M_1}_\a R^\a ) \quad ,
  \label{defalgebramoregeneral}
  \end{equation}
where $\Theta^{M_1}_\a$ is the embedding tensor corresponding to the
gauging of the internal symmetry. Considering together the analysis
of \cite{hierarchyE11} and the one performed in this section, we
write more generally the commutator of the 2-form generator with
momentum as
  \begin{equation}
  [ R^{\m_1 \m_2 , M_2 } , P_\n ] = ( W^{M_2}{}_{M_1} + \Theta^{N_1} S^{M_2}{}_{N_1 M_1} )
  \delta^{[\m_1}_\n R^{\m_2 ] , M_1} \quad ,
  \label{2formmomemtumgeneral}
  \end{equation}
where $W^{M_2}{}_{M_1}$ is related to the embedding tensor
$\Theta^{M_1}_\a$ as explained in \cite{hierarchyE11}. One then
derives the quadratic constraints of \cite{henning} from Jacobi
identities. In particular, using eqs. (\ref{defalgebramoregeneral})
and (\ref{2formmomemtumgeneral}) the Jacobi identity involving the
2-form and two momentum operators gives
  \begin{eqnarray}
  & & W^{M_2}{}_{M_1} \Theta^{M_1} + \Theta^{M_1} \Theta^{N_1}
  S^{M_2}{}_{N_1 M_1} = 0 \nonumber \\
  & & W^{M_2}{}_{M_1} \Theta^{M_1}_\a + a W^{M_2}{}_{M_1}
  \Theta^{P_1} D_{\a , P_1}{}^{M_1} + a \Theta^{N_1} \Theta^{P_1}
  S^{M_2}{}_{N_1 M_1} D_{\a , P_1 }{}^{M_1} \nonumber \\
  & & \quad \qquad + \Theta^{N_1}
  \Theta^{M_1}_\a S^{M_2}{}_{N_1 M_1} =0
   \quad ,
  \end{eqnarray}
while the Jacobi identity involving the 1-form generator, momentum
and the scalar generator $\Theta^{M_1 } K + a \Theta^{N_1} D_{\a ,
N_1 }{}^{M_1} R^\a + \Theta^{M_1}_\a R^\a$ gives
  \begin{eqnarray}
  & & \Theta^{N_1} \Theta^{Q_1} [ \delta^{M_1}_{N_1} \delta^{P_1}_{Q_1}
  + a D_{\alpha, N_1}{}^{M_1} D^\alpha_{Q_1}{}^{P_1} ] + \Theta^{M_1}_\a \Theta^{Q_1} D^\a_{Q_1}{}^{P_1} = 0
  \nonumber \\
  & & a \Theta^{N_1} \Theta^{P_1}_\a D_{\b , N_1}{}^{M_1} f^{\a
  \b}{}_\g - \Theta^{M_1}_\a \Theta^{P_1}_\b f^{\a\b}{}_\g + a^2
  \Theta^{N_1} \Theta^{Q_1} D_{\a , N_1}{}^{M_1} ( D_\a D^\g
  )_{Q_1}{}^{P_1} \nonumber \\
  & & \quad \  +a \Theta^{N_1} \Theta^{Q_1}_\g D_{\a ,
  N_1}{}^{M_1} D^\a_{Q_1}{}^{P_1} + a \Theta^{M_1}_\a \Theta^{Q_1}
  (D_\a D^\g )_{Q_1}{}^{P_1} - \Theta^{M_1}_\a \Theta^{Q_1}_\g
  D^\a_{Q_1}{}^{P_1} = 0 \quad .
  \label{parameterageneral}
  \end{eqnarray}
One can show that these quadratic constraints exactly reproduce the
quadratic constraints in \cite{henning} in all cases. In the rest of
this paper we will only consider for simplicity of notation the
embedding tensor $\Theta^{M_1}$ discussed so far in this section.
The analysis can be straightforwardly generalised to the case in
which both embedding tensors are turned on but it is technically
more involved.

The general method to compute the field strengths and gauge
transformations of the fields starting from the group element
  \begin{equation}
  g = e^{x \cdot P} e^{\Phi_{\rm Og} K^{\rm Og}}.... e^{A_{\m_1 ...\m_n , M_n} R^{\m_1 ... \m_n
  , M_n}} .., e^{A_{\m , M_1} R^{\m , M_1} } e^{\phi_\a R^\a}
  e^{h_\m{}^\n K^\m{}_\n}  \label{generalgroupelement}
  \end{equation}
is analogous to the one of \cite{hierarchyE11}. Using the algebraic
relations of eq. (\ref{defalgebra1}), (\ref{defalgebra2}) and
(\ref{defalgebra3}) one obtains from the Maurer-Cartan form the
field strengths
  \begin{eqnarray}
  & &  F_{\m_1 \m_2 , M_1} = 2 [ \de_{[\m_1} A_{\m_2 ] , M_1}
  -\frac{1}{2} \Theta^{Q_1} A_{[ \m_1 , N_1} A_{\m_2 ] , P_1} (
  \delta_{Q_1}^{N_1} \delta^{P_1}_{M_1} + a D_{\a , Q_1}{}^{N_1}
  D^\a_{M_1}{}^{P_1} ) \nonumber \\
  & & \quad \qquad - \Theta^{N_1} S^{M_2}{}_{N_1 M_1} A_{\m_1
  \m_2 , M_2} ]\nonumber \\
  & & F_{\m_1 \m_2 \m_3 , M_2} = 3 [ \de_{[\m_1} A_{\m_2 \m_3 ],
  M_2} + \frac{1}{2} \de_{[\m_1} A_{\m_2 , M_1} A_{\m_3 ] , N_1}
  f^{M_1 N_1}{}_{M_2} - A_{\m_1 \m_2 \m_3 , M_3} \Theta^{N_1}
  S^{M_3}{}_{N_1 M_2} \nonumber \\
  & & \quad \qquad - A_{[\m_1 \m_2 , N_2} A_{\m_3 ] , M_1}
  \Theta^{N_1} S^{N_2}{}_{N_1 P_1} f^{P_1 M_1}{}_{M_2} \nonumber \\
  & & \quad \qquad - \frac{a}{6}   A_{[ \m_1 , M_1 } A_{\m_2 , N_1} A_{\m_3 ] , P_1} \Theta^{Q_1} D_{\a , Q_1}{}^{M_1 }
  D^\a_{R_1}{}^{N_1}  f^{R_1 P_1}{}_{M_2} ] \nonumber \\
  & & F_{\m_1 \m_2 \m_3 \m_4 , M_3} = 4 [ \de_{[\m_1} A_{\m_2 \m_3
  \m_4 ] , M_3} - \de_{[\m_1} A_{\m_2 \m_3 , M_2} A_{\m_4 ] , M_1 }
  f^{M_1 M_2 }{}_{M_3} \nonumber \\
  & & \quad \qquad - \frac{1}{6} \de_{[\m_1} A_{\m_2 , M_1}
  A_{\m_3 , N_1} A_{\m_4 ], P_1} f^{M_1 N_1}{}_{M_2 } f^{P_1
  M_2}{}_{M_3} - A_{\m_1 \m_2 \m_3 \m_4 , M_4} \Theta^{N_1} S^{M_4
  }{}_{M_1 M_3}\nonumber \\
  &  &\quad \qquad  + A_{[ \m_1 \m_2 \m_3 , N_3} A_{\m_4 ] , M_1}
  \Theta^{N_1} S^{N_3}{}_{N_1 M_2} f^{M_1 M_2}{}_{M_3} \nonumber \\
  & & \quad \qquad - \frac{1}{2}
  A_{[\m_1 \m_2 , M_2} A_{\m_3 \m_4 ] , N_2} \Theta^{M_1}
  S^{M_2}{}_{M_1 N_1} f^{N_1 N_2}{}_{M_3} \nonumber \\
   &  &\quad \qquad
  + \frac{1}{2} A_{[ \m_1 \m_2 , M_2 } A_{\m_3 ,
  M_1 } A_{\m_4 ] , N_1 } \Theta^{P_1} S^{M_2}{}_{P_1 Q_1} f^{Q_1
  M_1}{}_{N_2} f^{N_1 N_2}{}_{M_3} \nonumber \\
  & & \quad \qquad + \frac{a}{24} A_{[ \m_1 , M_1} A_{\m_2 , N_1}
  A_{\m_3 , P_1} A_{\m_4 ] , Q_1} \Theta^{R_1}  D_{\a , R_1}{}^{M_1} D^\a_{S_1}{}^{N_1}
  f^{S_1 P_1}{}_{M_2} f^{Q_1 M_2}{}_{M_3} ]
  \label{generalformoffieldstrengths}
  \end{eqnarray}
for the 1-form, the 2-form and the 3-form potentials. For simplicity
we will not consider forms of higher rank in this paper. Acting on
the group element of eq. (\ref{generalgroupelement}) we also
determine the gauge transformations of these fields to be (see
Section 2 and Appendix B of \cite{hierarchyE11} for a detailed
derivation)
  \begin{eqnarray}
  & & \delta A_{\m , M_1} = a_{\m , M_1} + \Lambda_{N_1}
  \Theta^{P_1} ( \delta^{N_1}_{P_1} \delta^{Q_1}_{M_1} + a D_{\a ,
  P_1}{}^{N_1} D^\a_{M_1}{}^{Q_1} ) A_{\m , Q_1} \nonumber \\
  & & \delta A_{\m_1 \m_2 , M_2} = a_{\m_1 \m_2 , M_2} + \frac{1}{2}
  a_{[\m_1 , M_1} A_{\m_2 ] , N_1} f^{M_1 N_1}{}_{M_2} \nonumber \\
  & & \quad \qquad + \Lambda_{N_1} \Theta^{P_1} ( 2 \delta^{N_1}_{P_1} \delta^{Q_2}_{M_2} + a D_{\a ,
  P_1}{}^{N_1} D^\a_{M_2}{}^{Q_2} ) A_{\m_1 \m_2 , Q_2} \nonumber \\
  & & \delta A_{\m_1 \m_2 \m_3 , M_3} = a_{\m_1 \m_2 \m_3 , M_3}
  + a_{[\m_1 , M_1} A_{\m_2 \m_3 ] , M_2} f^{M_1 M_2}{}_{M_3}
  \nonumber \\
  & & \quad \qquad + \frac{1}{6} a_{[\m_1 , M_1} A_{\m_2, N_1}
  A_{\m_3 ] , P_1} f^{M_1 N_1}{}_{N_2}  f^{P_1 N_2}{}_{M_3}
  \nonumber \\
  & & \quad \qquad  + \Lambda_{N_1} \Theta^{P_1} ( 3 \delta^{N_1}_{P_1} \delta^{Q_3}_{M_3} + a D_{\a ,
  P_1}{}^{N_1} D^\a_{M_3}{}^{Q_3} ) A_{\m_1 \m_2 \m_3 , Q_3} \quad ,
  \label{generalformofgaugetransfs}
  \end{eqnarray}
where the parameters $a_{\m_1 ... \m_n , M_n}$ are given in terms of
the gauge parameters $\Lambda_{\m_1 ...\m_{n-1} , M_n}$ as
  \begin{equation}
  a_{\m_1 ... \m_n , M_n} = \de_{[\m_1} \Lambda_{\m_2 ...\m_{n}] ,
  M_n} + \Theta^{M_1} S^{M_{n+1}}{}_{M_1 M_n } \Lambda_{\m_1 ...\m_n
  , M_{n+1}} \quad . \label{aintermsoflambda}
  \end{equation}

Using eqs. (\ref{purespinorconstraint}) and (\ref{parametera}) one
can show  that the embedding tensor projects $A_{\m , M_1}$ onto the
abelian vector transforming as \cite{henning}
  \begin{equation}
  \delta (\Theta^{M_1} A_{\m , M_1} ) = \de_{\m } ( \Theta^{M_1}
  \Lambda_{M_1} ) \quad .
  \end{equation}
What we are describing is thus an abelian gauging in a formulation
which is formally covariant under $E_{11-D}$. All the other vectors
are abelian vectors which are also charged with respect to the
gauged $U(1)$. Some of these vectors are gauged away by the
parameter $\Lambda_{\m , M_2}$, and the hierarchical structure
continues to higher rank forms exactly as in the case of the
internal gaugings.

We expect the dynamics to arise from duality relations among the
field strengths. This was indeed shown in three dimensions in
\cite{henning} by imposing the closure of the supersymmetry algebra.
In \cite{fabiopeterE11origin,ericembeddingtensor} it was shown that
decomposing the $E_{11}$ algebra with respect to $GL(D, \mathbb{R})
\otimes E_{11-D}$ one finds $D-1$ form generators in the same
$E_{11-D}$ representation as the embedding tensor $\Theta^{M_1}_\a$
associated to the internal gaugings. Correspondingly, one expects
the supersymmetry algebra to close on the corresponding fields,
provided that a duality relation between the field strength of the
$D-1$ form potential and the embedding tensor holds. This was indeed
explicitly shown first in the case of Romans IIA
\cite{diederikIIA,fabioericIIA}, and then in the five dimensional
\cite{E11fivedimensions} and three dimensional
\cite{hierarchythreedimensions} cases. In
\cite{erichalfmax,hierarchythreedimensions} it was also observed
that the $D$ form generators predicted by $E_{11}$ contain in all
cases the $E_{11-D}$ representations associated to the quadratic
constraints of the embedding tensor $\Theta^{M_1}_\a$.

The $E_{11}$ spectrum does not possess non-propagating forms (that
is forms of rank higher than $D-2$) in the same representation as
the embedding tensor $\Theta^{M_1}$, and thus this embedding tensor
associated to the trombone deformations discussed in this paper can
not be related by duality to an $E_{11}$ form. In \cite{henning} it
was conjectured that the generators in the representation $GL(D,
\mathbb{R})$ of mixed symmetry that we denote by $(1,D-2)$, that is
the generators $R^{\m, \n_1 ...\n_{D-2} , M_1}$, that are in the
same $E_{11-D}$ representation as the 1-form generators and thus as
the trombone embedding tensor, might trigger these deformations. As
the quadratic constraint of eq. (\ref{purespinorconstraint})
projects out the representation ${\bf R_2}$ to which the 2-form
generators belong, in \cite{henning} it was also conjectured that
the $E_{11}$ generators $R^{\m_1 \m_2, \n_1 ... \n_{D-2} , M_2}$ in
the $(2,D-2)$ representation of $GL(D, \mathbb{R})$ might be
associated to these quadratic constraints. Observe that the presence
of these generators in the $E_{11}$ spectrum is completely general,
as shown in \cite{E11dualfields}. Indeed these are the first of an
infinite chain of so called ``dual'' $n$-form generators in the
$GL(D, \mathbb{R})$ representations $(n , D-2 , D-2 ,..., D-2)$ and
in the same $E_{11-D}$ representation ${\bf R_n}$ as the $n$-form
generators. Their presence is crucial for the universal structure of
$E_{11}$ reproducing the gauge algebra of all the form fields in all
dimensions. In \cite{erichalfmax,hierarchythreedimensions} it was
observed that in the case of the internal gaugings one can consider
a lagrangian formulation in which the $D-1$ forms are Lagrange
multipliers for the embedding tensor (so that their field equation
implies the constancy of the embedding tensor) and the $D$ forms are
Lagrange multipliers for the quadratic constraints. The fact that
these forms are present in the gauge algebra is thus intrinsically
related to the fact that one expects such a lagrangian formulation
to be possible. The theories considered in this paper do not admit a
lagrangian formulation, because they result from the gauging of the
trombone symmetry, which is not a symmetry of the lagrangian but
only of the field equations. Thus we consider the fact that there
are no form generators associated to these deformations as
completely consistent, and we do not expect any $E_{11}$ generator
associated to a non-propagating field to play a role in triggering
these deformations.

The main difference with respect to the deformations of the
$E_{11,D}^{local}$ algebra studied in \cite{hierarchyE11} is that in
this case the deformation also involves the $GL(D, \mathbb{R})$
generators, and we thus have to consider the gravity sector as well.
The modification of the gravity part is precisely as explained in
the ten-dimensional gauged IIA case in \cite{trombone1}. Starting
from the group element of eq. (\ref{generalgroupelement}) and
considering the deformed commutator of eq. (\ref{defalgebra1}), the
$K^a{}_b$ term in the Maurer-Cartan form becomes
  \begin{equation}
  \left[ (e^{-1} \de_\m e )_{ab} - \Phi_{\m\n}^\r e^\n_a e_{\r b} -
  \Theta^{M_1} A_{\m , M_1} \eta_{ab} \right ] K^{ab} \quad , \label{gravitydeformed}
  \end{equation}
where $\Phi_{\m\n}^\r $ is the gravity Og 1 field. Observing that
the vierbein transforms under the gauge parameter of the 1-form
$\Lambda_{M_1}$ as
  \begin{equation}
  e_\m{}^a \rightarrow e^{\Theta^{M_1} \Lambda_{M_1}} e_{\m}{}^a \quad ,
  \label{vierbeintransf}
  \end{equation}
we write the term contracting $K^{ab}$ in eq.
(\ref{gravitydeformed}) as
  \begin{equation}
  (e^{-1} D_\m e )_{ab} - \Phi_{\m\n}^\r e^\n_a e_{\r b}
   \quad , \label{deformedspinconnectionchristoffel}
  \end{equation}
where $D_\m$ is the derivative covariantised with respect to the
transformation of eq. (\ref{vierbeintransf}), that is $D_\m = \de_\m
- \Theta^{M_1} A_{\m, M_1}$. Applying the same arguments of
\cite{fabiopeterogievetsky} we obtain that imposing that the
symmetric part in $ab$ of eq.
(\ref{deformedspinconnectionchristoffel}) vanishes gives for the
antisymmetric part the spin connection
  \begin{equation}
  \omega_{\mu}{}^{ab} ={1\over
  2} e^{\n a}( D_\mu e_\n{}^b - D_\n e_\mu{} ^b )
  - {1\over 2} e^{\n b }( D_\mu e_\n{}^a
  - D_\n e_\mu{}^a ) -{1\over 2} e^{ \n a } e^{ \r  b}
  ( D_\n e_\r{}^c - D_\r
  e_\n{}^c )e_\mu{}^c  \quad , \label{spinconnection}
  \end{equation}
that is
 \begin{equation}
 \tilde{\omega}_\m{}^{ab} = \omega_\m{}^{ab} - 2 \Theta^{M_1} e_\m{}^{[a} e^{| \n
 | b]} A_{\n , M_1} \quad . \label{deformedspinconnection}
 \end{equation}
If one plugs this into the Maurer-Cartan form and applies the
inverse Higgs mechanism at the level of the next gravity Og field,
one obtains that the term contracting $K^{ab}{}_c$ is the
covariantised Riemann tensor
  \begin{equation}
  \tilde{R}_{\m\n}{}^{ab} = 2 \de_{[\m} \tilde{\omega}_{\n ]}{}^{ab} + 2
  \tilde{\omega}_{[\m}^{ac} \tilde{\omega}_{\n ] c}{}^b \quad .
  \label{deformedriemanntensor}
  \end{equation}
This reproduces exactly the analysis of \cite{henning} in the
gravity sector.

The rest of this paper will be devoted to a careful analysis of the
trombone deformations in each dimensions.

\section{D=3}
The massless maximal supergravity theory in three dimensions was
constructed in \cite{D=3masslesssugra}. Its bosonic sector describes
128 scalars parametrising the manifold $E_{8(+8)}/SO(16)$ and the
metric. This theory arises from the $E_{11}$ decomposition
appropriate to three dimensions, corresponding to the deletion of
node 3 in the Dynkin diagram of fig. 1. We consider the
decomposition of the adjoint of $E_{11}$ in terms of $GL(3,
\mathbb{R}) \otimes E_{8 (8)}$ and we only take into account the
level zero generators and the positive level generators which are
forms of rank at most two, which are
   \begin{equation}
   K^\m{}_\n \ ({\bf 1}) \quad R^{\alpha} \  ({\bf {248}}) \quad R^{\m , \alpha}  \ ({\bf 248} )
  \quad R^{\m_1 \m_2, M}  \ ( {\bf 3875} ) \quad R^{\m_1 \m_2}
   \   ({\bf 1} )
  \quad . \label{generatorsinthreedimensions}
  \end{equation}
The 1-form generators are in the same $E_8$ representation of the
scalars, corresponding to the fact that in three dimensions vectors
are dual to scalars, while the 2-form generators are in the ${\bf 1
\oplus 3875}$.

The commutators of these generators which are relevant for our
purpose were given in \cite{hierarchyE11}, and are
  \begin{eqnarray}
  & & [ R^\alpha , R^\beta ] = f^{\alpha \beta}{}_\gamma R^\gamma
  \nonumber \\
  & & [ R^\alpha , R^{\m , \beta} ] = f^{\alpha \beta}{}_\gamma R^{ \m,
  \gamma}
  \nonumber \\
  & & [ R^\alpha , R^{\m_1 \m_2 } ] = 0 \nonumber \\
  & & [ R^\alpha , R^{\m_1 \m_2 , M} ] = D^\alpha_N{}^M R^{\m_1 \m_2 ,
  N} \nonumber \\
  & & [ R^{\m_1 , \alpha} , R^{\m_2 ,\beta} ] = g^{\alpha \beta}
  R^{\m_1 \m_2} + S^{\alpha \beta}_M R^{\m_1 \m_2 , M} \quad .
  \label{3dimalgebramassless}
  \end{eqnarray}
We take our $E_8$ conventions from \cite{E8conventions}, and thus
$g^{\alpha \beta}$ is the Cartan-Killing metric, defined by
  \begin{equation}
  g^{\alpha \beta} = -{1 \over 60} f^{\alpha}{}_{\gamma \delta}
  f^{\beta \gamma \delta} \quad ,
  \end{equation}
and it is the metric we use to raise $E_8$ indices in the adjoint,
$D^\alpha_N{}^M$ are the $E_8$ generators in the ${\bf 3875}$ and
$S^{\alpha \beta}_M$ is an $E_8$ invariant tensor symmetric in
$\alpha \beta$.  The symmetric product of two adjoint
representations of $E_8$ is
  \begin{equation}
  {\bf [ 248 \otimes 248 ]_{\rm S} = 1 \oplus 3875 \oplus 27000 }
  \quad , \label{248symmproduct}
  \end{equation}
and we will need the $E_8$ projectors
  \begin{eqnarray}
  & & \mathbb{P}_{\bf 1}{}_{\alpha \beta}{}^{\gamma \delta} =
  \frac{1}{248} g_{\alpha \beta} g^{\gamma \delta} \nonumber \\
  & & \mathbb{P}_{\bf 3875}{}_{\alpha \beta}{}^{\gamma \delta} =
  \frac{1}{7} \delta^{\gamma}_{(\alpha} \delta^{\delta}_{\beta )} -
  \frac{1}{14} f^{\sigma \gamma}{}_{(\alpha} f_{\beta )
  \sigma}{}^{\delta} - \frac{1}{56} g_{\alpha \beta} g^{\gamma
  \delta} \nonumber \\
  & & \mathbb{P}_{\bf 27000}{}_{\alpha \beta}{}^{\gamma \delta} =
  \frac{6}{7} \delta^{\gamma}_{(\alpha} \delta^{\delta}_{\beta )} +
  \frac{1}{14} f^{\sigma \gamma}{}_{(\alpha} f_{\beta )
  \sigma}{}^{\delta} + \frac{3}{217} g_{\alpha \beta} g^{\gamma
  \delta}
  \quad , \label{E8projectors}
  \end{eqnarray}
which project the symmetric product of two adjoint indices $\alpha
\beta$ into the singlet, the ${\bf 3875}$ and the ${\bf 27000}$
respectively. The invariant tensor $S^{\alpha \beta}_M$ must satisfy
constraints that project the symmetric indices $\alpha \beta$ on the
${\bf 3875}$, and from eq. (\ref{E8projectors}) one gets
\cite{hierarchyE11}
  \begin{eqnarray}
  & & g_{\alpha \beta} S^{\alpha \beta}_M = 0 \nonumber \\
  & & S^{\alpha \beta}_M = - {1 \over 12}
  f^\epsilon{}_\gamma{}^\alpha f_{\epsilon \delta}{}^\beta
  S^{\gamma \delta}_M \quad . \label{conditionsonSinthree}
  \end{eqnarray}

Following the general analysis of section 2, we now consider the
trombone deformation of this algebra. We start form the deformed
commutator
  \begin{equation}
  [ R^{\m , \alpha} , P_\n ] =  \delta^\m_\n ( \Theta^\alpha K +a
  f^{\alpha}{}_{\beta \gamma} \Theta^\beta R^\gamma ) \quad ,
  \end{equation}
with the parameter $a$ to be determined. We also write the
commutators of the 2-form generators with momentum,
   \begin{eqnarray}
   & & [ R^{\m_1 \m_2 } , P_\n ] = b  \Theta^\alpha \delta^{[\m_1}_\n
   R^{\m_2 ] , \beta} g_{\alpha \beta} \nonumber \\
   & & S^{\gamma \delta}_M [ R^{\m_1 \m_2 , M} , P_\n ] = c
   \Theta^\alpha \delta^{[\m_1}_\n
   R^{\m_2 ] , \beta} \mathbb{P}_{\bf 3875}{}_{\alpha \beta}{}^{\gamma \delta}
   \end{eqnarray}
with additional parameters $b$ and $c$ to be determined.

We first determine the quadratic constraints resulting from imposing
the closure of the Jacobi identities involving the 2-form generator
and two momentum operators. This gives the $E_8$ case of eq.
(\ref{purespinorconstraint}), that is
  \begin{eqnarray}
  & & g_{\alpha \beta} \Theta^\alpha \Theta^\beta =0 \nonumber \\
  & & \mathbb{P}_{\bf 3875}{}_{\alpha \beta}{}^{\gamma \delta} \Theta^\alpha
  \Theta^\beta =0 \quad . \label{projectoutE8}
  \end{eqnarray}
We then impose the closure of the Jacobi identity involving $R^{\m ,
\a}$, $R^{\n , \b}$ and $P_\r$, obtaining the condition
  \begin{equation}
  \left( 2 - \frac{c}{7} \right) \Theta^{(\a} R^{\m , \b )} + \left( 2 a + \frac{c}{14} \right)\Theta^\g f^{(\a}{}_{\g \d} f
  ^{| \d | \b)}{}_{\e} R^{\m , \e } + \left( \frac{c}{56} - b \right)  g^{\a \b} g_{\g\d} \Theta^\g
  R^{\m , \d} = 0 \quad ,
  \end{equation}
which implies
  \begin{equation}
  a = -\frac{1}{2} \qquad \quad b = \frac{1}{4} \quad \qquad c = 14
  \quad .
  \end{equation}
Remarkably, for this value of $a$ the Jacobi identity between the
scalar operator $\Theta^\alpha K + a f^\a{}_{\b \g} \Theta^\b R^\g$,
the 1-form generator $R^{\m , \a}$ and momentum, which is eq.
(\ref{parametera}), becomes
  \begin{equation}
  \Theta^\a \Theta^\b - \frac{1}{2} f^\a{}_{\g\d} f^{\d\b}{}_\e
  \Theta^\g \Theta^\e = 0 \quad .
  \end{equation}
This is exactly the quadratic constraint derived in the field theory
analysis of \cite{henning}. As it is clear from the projector
operators of eq. (\ref{E8projectors}), this equation projects the
product of two embedding tensors on the ${\bf 27000}$, and given eq.
(\ref{248symmproduct}), this condition is equivalent to eq.
(\ref{projectoutE8}), which projects out the singlet and the ${\bf
3875}$. This proves in this $E_8$ case the statement made in section
2 that the condition resulting from this Jacobi identity is already
contained in the ``pure spinor constraint'' of eq.
(\ref{purespinorconstraint}).

To summarise, we have obtained the commutators
  \begin{eqnarray}
  & & [ R^{\m ,\a } , P_\n ] = \delta^\m_\n ( \Theta^\a K -
  \frac{1}{2} f^\a{}_{\b\g} \Theta^\b R^\g ) \nonumber \\
  & & [ R^{\m_1 \m_2 } , P_\n ] = \frac{1}{4}   g_{\a\b} \Theta^\a
  \delta^{[\m_1}_\n R^{\m_2 ] , \b } \nonumber \\
  & &   S^{\gamma \delta}_M [ R^{\m_1 \m_2 , M} , P_\n ] = 14
   \Theta^\alpha \delta^{[\m_1}_\n
   R^{\m_2 ] , \beta} \mathbb{P}_{\bf 3875}{}_{\alpha \beta}{}^{\gamma
   \delta} \quad , \label{summaryD=3trombone}
   \end{eqnarray}
and we normalise the invariant tensor $S^{\a\b}_M$ in such a way
that
  \begin{equation}
  S^{\a\b}_M S_{\a\b , N} = \delta_{MN} \quad ,
  \end{equation}
so that the last of eqs. (\ref{summaryD=3trombone}) can be written
 \begin{equation}
   [ R^{\m_1 \m_2 , M} , P_\n ] = 14
   \Theta^\alpha \delta^{[\m_1}_\n
   R^{\m_2 ] , \beta} S_{\a\b}^{ M} \quad .
   \end{equation}

 To conclude this section, we apply the general formulae of
section 2 to derive the field strengths and gauge transformations of
the fields. We consider the group element
  \begin{equation}
  g =  e^{x \cdot P} e^{A_{\m_1 \m_2} R^{\m_1 \m_2}} e^{A_{\m_1 \m_2 ,M} R^{\m_1 \m_2
  ,M}} e^{A_{\m , \alpha} R^{\m , \alpha}} e^{\phi_{\alpha}
  R^{\alpha}} e^{h_\m{}^\n K^\m{}_\n}
  \label{groupelementD=3} \quad ,
  \end{equation}
and we compute the Maurer-Cartan form and the gauge transformations
using the commutators of eq. (\ref{3dimalgebramassless}) and
(\ref{summaryD=3trombone}). This gives the field strength for the
1-form
  \begin{eqnarray}
  & & F_{\m_1 \m_2 , \a} = 2 [ \de_{[\m_1} A_{\m_2 ] , \a} -\frac{1}{2}
  \Theta^\b A_{[\m_1 , \b } A_{\m_2 ], \a} + \frac{1}{4} \Theta^\g
  f^\b{}_{\g\d}f^{\d \e}{}_\a A_{[\m_1 , \b } A_{\m_2 ] , \e} \nonumber \\
  & & \quad  -
  \frac{1}{4} \Theta_\a A_{\m_1 \m_2} -14 \Theta^\b S^M_{\a\b}
  A_{\m_1 \m_2 , M}] \quad ,
  \end{eqnarray}
transforming covariantly under the gauge transformations
  \begin{eqnarray}
  & & \d A_{\m , \a} = a_{\m , \a} + \Lambda_\b \Theta^\b A_{\m ,
  \a} - \frac{1}{2} \Lambda_\d \Theta^\b f^\d{}_{\b\g} f^{\g\s}{}_\a
  A_{\m ,\s} \nonumber \\
  & & \d A_{\m_1 \m_2} = \de_{[\m_1} \Lambda_{\m_2 ] } + \frac{1}{2}
  g^{\a\b} a_{[\m_1 , \a} A_{\m_2 ] , \b} + 2 \Lambda_\a \Theta^\a
  A_{\m_1 \m_2 } \nonumber \\
  & & \d A_{\m_1 \m_2 , M} = \de_{[\m_1} \Lambda_{\m_2 ] ,M} +
  \frac{1}{2} S^{\a\b}_M a_{[\m_1 , \a} A_{\m_2 ] , \b} + 2
  \Lambda_\a \Theta^\a A_{\m_1 \m_2 , M} \nonumber \\
  & & \quad - \frac{1}{2} \Lambda_\a \Theta^\b f^\a{}_{\b\g}
  D^\g_M{}^N A_{\m_1 \m_2 , N} \quad ,
 \end{eqnarray}
where $D^\a_M{}^N$ are the $E_8$ generators in the ${\bf 3875}$ and
  \begin{equation}
  a_{\m , \a} = \de_\m \Lambda_\a  + \frac{1}{4} \Theta_\a
  \Lambda_\m + 14 \Theta^\b S^M_{\a\b} \Lambda_{\m , M} \quad .
  \end{equation}

We can also compute the field strengths of the 2-form potentials up
to terms involving the 3-form potential. The result is
  \begin{eqnarray}
  & & F_{\m_1 \m_2 \m_3} = 3 [ \de_{[\m_1} A_{\m_2 \m_3 ]} +
  \frac{1}{2} g^{\a\b} \de_{[\m_1} A_{\m_2 , \a} A_{\m_3 ],\b} -
  \frac{1}{4} \Theta^\a A_{[\m_1 \m_2 } A_{\m_3 ] ,\a} \nonumber \\
  & & \quad -14 \Theta^\a
  S^M_{\a\b} A_{[\m_1 \m_2 , M} A_{\m_3 ]}^\b +
  \frac{1}{ 12} A_{[\m_1
  , \a} A_{\m_2 , \b} A_{\m_3 ] , \g }  \Theta^\d f^\a{}_{\d\s}
  f^{\s\b\g}] \nonumber \\
  & & F_{\m_1 \m_2 \m_3, M} = 3 [ \de_{[\m_1} A_{\m_2 \m_3 ], M} +
  \frac{1}{2} S^{\a\b}_M \de_{[\m_1} A_{\m_2 , \a} A_{\m_3 ],\b} -
  \frac{1}{4} A_{[\m_1 \m_2} A_{\m_3 ] , \a} \Theta_\b S^{\a\b}_M
  \nonumber \\
  & & \quad - 14 A_{[\m_1 \m_2 , N} A_{\m_3 ] , \a} \Theta^\b
  S^N_{\b\g} S^{\a\g}_M +
  \frac{1}{ 12} A_{[\m_1
  , \a} A_{\m_2 , \b} A_{\m_3 ] , \g }  \Theta^\d f^\a{}_{\d\s}
  f^{\s\b}{}_\e S^{\e\g}_M ] \quad .
  \end{eqnarray}
To prove the gauge covariance of these field strengths of the
2-forms one must include the 3-forms and determine their gauge
transformations.

\section{D=4}
The deletion of node 4 in the $E_{11}$ Dynkin diagram of fig. 1
results in a decomposition of $E_{11}$ with respect to $GL(4,
\mathbb{R}) \otimes E_{7(7)}$, which is relevant for the
four-dimensional theory. The massless maximal supergravity theory
was constructed in \cite{cremmerjuliaD=4}, and possesses an
$E_{7(7)}$ on-shell symmetry (recently, it was shown that this
theory also admits a lagrangian formulation in which the $E_{7(7)}$
symmetry is manifest \cite{christian}). The general gaugings of the
internal symmetry $E_{7(7)}$ of this theory were constructed in
\cite{dWSTD=4} using the embedding tensor formalism, and in
\cite{hierarchyE11} it was shown that all these gaugings correspond
to deformations of the four-dimensional local $E_{11}$ algebra.

We use the notation of \cite{hierarchyE11}, which we now review. We
consider the level zero generators and the form generators, which
are
    \begin{eqnarray}
   & & K^\m{}_\n \ ({\bf 1}) \ \   R^{\alpha} \ ({\bf {133}}) \  \  R^{\m , M}  \ ({\bf 56} )
  \  \ R^{\m_1 \m_2, \alpha}  \ ( {\bf 133} ) \nonumber \\
  & &    R^{\m_1 \m_2 \m_3 , A}
   \   ({\bf 912} ) \ \  R^{\m_1 ...\m_4 , \alpha \beta} \ ({\bf 8645
   \oplus 133})
  \quad , \label{genlistD=4}
  \end{eqnarray}
where the numbers in brackets denote the corresponding $E_7$
representation. The indices $M$, denoting the ${\bf 56}$, are raised
and lowered by the antisymmetric invariant metric $\Omega^{MN}$
according to
  \begin{equation}
  V^M = \Omega^{MN} V_N \qquad V_M = V^N \Omega_{NM} \quad ,
  \label{raiseandlowerinfourdim}
  \end{equation}
which implies
  \begin{equation}
  \Omega^{MN} \Omega_{NP} = - \delta^M_P \quad .
  \end{equation}
The generator
  \begin{equation}
  D^{\alpha , MN} = \Omega^{MP} D^\alpha_P{}^N
  \end{equation}
is symmetric in $MN$.

The commutators between the scalars and the other generators are
dictated by the $E_7$ representation that the generators carry. The
other $E_{11}$ commutation relations which will be relevant in this
paper are \cite{hierarchyE11}
  \begin{eqnarray}
  & & [ R^{\m_1 , M} , R^{\m_2 , N} ] = D_\alpha^{MN} R^{\m_1 \m_2 ,
  \alpha} \nonumber \\
  & & [ R^{\m_1 , M} , R^{\m_2 \m_3 , \alpha} ] = S^{M \alpha}_A R^{\m_1
  \m_2 \m_3 , A} \nonumber \\
  & & [ R^{\m_1 \m_2 , \alpha} , R^{\m_3 \m_4 , \beta} ] = R^{\m_1 ...
  \m_4, \alpha \beta} \nonumber \\
  & & [ R^{\m_1 , M} , R^{\m_2 \m_3 \m_4 , A} ] = C^{MA}_{\alpha \beta}
  R^{\m_1 ... \m_4 , \alpha \beta}
  \label{E11algebrainfourdimensions}
  \quad ,
  \end{eqnarray}
and the Jacobi identities as well as the condition of invariance
under $E_7$ imply that $S^{M \alpha}_A$ satisfies
\cite{hierarchyE11}
    \begin{equation}
  D_\alpha^{(MN} S^{P) \alpha}_A = 0
  \label{otherconditiononS}
  \end{equation}
and
  \begin{equation}
  D_{\alpha , M}{}^N S^{M \alpha}_A = 0 \quad .
  \label{gravitinoconditiononS}
  \end{equation}
These conditions project the $M\alpha$ indices of $S^{M \alpha}_A$
along the ${\bf 912}$, as can be seen applying the projector
operator
  \begin{equation}
  \mathbb{P}_{\bf 912}{}_{\alpha N}^{M \beta} = {4 \over 7} D^\beta_N{}^P D_{\alpha
  P}{}^M - {12 \over 7}  D_{\alpha N}{}^P D^\beta_{P}{}^M + {1 \over 7 } \delta^M_N
  \delta^\beta_\alpha \quad . \label{912projector}
  \end{equation}
Following \cite{dWSTgeneral}, we are using the metric
  \begin{equation}
  g^{\alpha \beta} = D^\alpha_M{}^N D^\beta_N{}^M
  \label{E7metricproptokilling}
  \end{equation}
to raise and lower indices in the adjoint. This metric is
proportional to the Cartan-Killing metric, as can be seen from
  \begin{equation}
  f_{\alpha \beta\gamma} f^{\alpha \beta \delta} = - 3
  \delta^\delta_\gamma
  \quad .
  \end{equation}
The other invariant tensor $C^{MA}_{\alpha \beta}$ satisfies
    \begin{equation}
  S^{M \alpha}_A C^{NA}_{\beta \gamma} + S^{N \alpha}_A C^{MA}_{\beta
  \gamma} + \delta^\alpha_{[\beta} D_{\gamma ]}^{MN} = 0 \quad .
  \end{equation}
Including also the momentum operator and the Og operators one
constructs the algebra $E_{11,4}^{local}$ from which one derives the
field strengths and gauge transformations of all the fields of the
massless maximal supergravity theory in four dimensions
\cite{hierarchyE11}.

We now consider the trombone deformation of the $E_{11,4}^{local}$
algebra. We consider as our starting point the commutator
  \begin{equation}
  [ R^{\m , M} , P_\n ] = \delta^\m_\n ( \Theta^M K + a \Theta^N
  D_{\alpha, N}{}^M R^\alpha ) \label{startingdefD=4}
  \quad .
  \end{equation}
The Jacobi identity among $R^{\m , M}$, $R^{\n , N}$ and $P_\r$
implies that
  \begin{equation}
  [ R^{\m_1 \m_2 ,\alpha} , P_\n ] = -16 \Theta^M D^\alpha_{MN}
  \delta^{[\m_1}_\n R^{\m_2 ] , N} \quad ,
  \end{equation}
and also fixes the parameter $a$ in eq. (\ref{startingdefD=4}) to be
  \begin{equation}
  a = -8 \quad . \label{parameteraintheE7case}
  \end{equation}
The Jacobi identity among $R^{\m_1 , M}$, $R^{\m_2 \m_3 , \alpha}$
and $P_\n$ then gives
  \begin{equation}
  S^{M \alpha}_A [ R^{\m_1 \m_2 \m_3 , A} , P_\n ] = 14  \mathbb{P}_{\bf 912}{}_{\gamma N}^{M \delta}
  g^{\a\g} g_{\b\d} \Theta^N \delta^{[\m_1 }_\n R^{\m_2 \m_3 ] ,
  \beta} \quad .
  \end{equation}
In deriving these results we have made use of the $E_7$ identity
  \begin{equation}
  D_{\beta M}{}^N D^\beta_P{}^Q = {1 \over 12} \delta_M^Q \delta^N_P
  + {1 \over 24} \delta_M^N \delta_P^Q - {1 \over 24} \Omega^{NQ}
  \Omega_{MP} + D_{\beta}^{NQ} D^\beta_{MP} \quad .
  \label{E7identity}
  \end{equation}
The $E_{11}$-based derivation of the $E_7$ relations of eqs.
(\ref{912projector}) and (\ref{E7identity}) can be found in Appendix
A of ref. \cite{hierarchyE11}. The Jacobi identity involving two
2-form generators and momentum gives
  \begin{equation}
  [ R^{\m_1 ...\m_4 , \a \b} , P_\n ] = -32 \Theta^M D^{[\a}_{MN}
  S^{\b ] N}_A \delta^{[\m_1}_\n R^{\m_2 \m_3 \m_4 ] , A} \quad ,
  \end{equation}
and one can show that the Jacobi identity involving the 1-form, the
3-form and momentum is also automatically satisfied.

We now consider the quadratic constraints. The Jacobi identity
involving $R^{\m_1 \m_2 , \a}$ and two momentum operators gives the
$E_7$ case of the ``pure spinor condition'' of eq.
(\ref{purespinorconstraint}), which is
  \begin{equation}
  \Theta^M \Theta^N D^\a_{MN} =0 \quad ,
  \label{purespinorconstraintE7}
  \end{equation}
which again coincides with the quadratic constraint of
\cite{henning}. Using the $E_7$ relations of Appendix A of ref.
\cite{hierarchyE11}, some of which we have reviewed in this paper,
one can show that this constraint automatically implies that the
Jacobi identity involving $R^{\m_1 \m_2 \m_3 , A}$ and two momentum
operators is automatically satisfied.

Finally, we consider the Jacobi identity among the scalar generator
$\Theta^M K + a \Theta^N D_{\a , N}{}^M R^\a$, the 1-form $R^{\m ,
M}$ and $P_\n$. A simple computation shows that for the value $a= -8
$ as in eq. (\ref{parameteraintheE7case}) already determined
imposing the closure of other Jacobi identities, this Jacobi
identity is automatically satisfied if the ``pure spinor condition''
of eq. (\ref{purespinorconstraintE7}) holds.

To summarise, we have obtained the commutators
  \begin{eqnarray}
  & &   [ R^{\m , M} , P_\n ] = \delta^\m_\n ( \Theta^M K -8 \Theta^N
  D_{\alpha, N}{}^M R^\alpha ) \nonumber \\
  & &   [ R^{\m_1 \m_2 ,\alpha} , P_\n ] = -16 \Theta^M D^\alpha_{MN}
  \delta^{[\m_1}_\n R^{\m_2 ] , N} \nonumber \\
  & &    S^{M \alpha}_A [ R^{\m_1 \m_2 \m_3 , A} , P_\n ] = 14  \mathbb{P}_{\bf 912}{}_{\gamma N}^{M \delta}
  g^{\a\g} g_{\b\d} \Theta^N \delta^{[\m_1 }_\n R^{\m_2 \m_3 ] ,
  \beta}  \nonumber \\
  & &   [ R^{\m_1 ...\m_4 , \a \b} , P_\n ] = -32 \Theta^M D^{[\a}_{MN}
  S^{\b ] N}_A \delta^{[\m_1}_\n R^{\m_2 \m_3 \m_4 ] , A}
  \quad . \label{deformedalgebrainfour}
  \end{eqnarray}
Normalising the invariant tensor $ S^{M \alpha}_A$ in such a way
that
  \begin{equation}
  S^{M\alpha}_A S_{M \alpha}^B = \d^B_A \quad ,
  \end{equation}
the third equation in (\ref{deformedalgebrainfour}) can be rewritten
as
  \begin{equation}
  [ R^{\m_1 \m_2 \m_3 , A} , P_\n ] = 14 \Theta^M S^A_{M\a} \delta^{[\m_1 }_\n R^{\m_2 \m_3 ] ,
  \a} \quad .
  \end{equation}

We now consider the group element
\begin{equation}
  g =  e^{x \cdot P} e^{A_{\m_1
  ...\m_4, \alpha\beta} R^{\m_1 ... \m_4, \alpha\beta}} ...
  e^{A_{\m
  , M} R^{\m , M}} e^{\phi_{\alpha} R^{\alpha}} e^{h_\m{}^\n K^\m{}_\n } \quad ,
  \label{fourdimensionalgroupelement}
  \end{equation}
and applying  the general formulae of section 2 to derive the field
strengths and gauge transformations of the fields, we get
  \begin{eqnarray}
  & & F_{\m_1 \m_2 , M} = 2 [ \de_{[\m_1} A_{\m_2 ] , M} +
  \frac{1}{6} \Theta^N A_{[\m_1 , M } A_{\m_2 ], N} -\frac{1}{6}
  \Theta_M \Omega^{NP} A_{[\m_1 ,N} A_{\m_2 ] , P} \nonumber \\
   & & \quad + 16 \Theta^N
  D^\a_{MN} A_{\m_1 \m_2 , \a} ] \nonumber \\
  & & F_{\m_1 \m_2 \m_3 , \a} = 3 [ \de_{[\m_1} A_{\m_2 \m_3 ] , \a}
  + \frac{1}{2} \de_{[\m_1} A_{\m_2 , M} A_{\m_3 ] , N  } D_\a^{MN}
  - 14 \Theta^M S^A_{M\a} A_{\m_1 \m_2 \m_3 ,A} \nonumber \\
  & & \quad  + 16 A_{[\m_1 \m_2 ,
  \b} A_{\m_3 ] , M} \Theta^N D^\b_{NP}  D_\a^{MP} + \frac{4}{3}
  A_{[\m_1 , M } A_{\m_2 , N} A_{\m_3 ] , P} \Theta^Q D_{\b , Q}{}^M
  D^\b_R{}^N D_\a^{RP} ] \nonumber \\
  & & F_{\m_1 \m_2 \m_3 \m_4 , A} = 4 [ \de_{[\m_1} A_{\m_2 \m_3
  \m_4 ] , A} - \de_{[\m_1} A_{\m_2 \m_3 , \a} A_{\m_4 ] , M}
  S^{M\a}_A \nonumber \\
  & & \quad - \frac{1}{6} \de_{[\m_1} A_{\m_2  , M} A_{\m_3 , N }
  A_{\m_4 ] , P} D_\a^{MN} S^{P\a}_A + 32 \Theta^M D^{[\a}_{MN}
  S^{\b ] N}_A A_{\m_1 ...\m_4 , \a\b} \nonumber \\
  & & \quad + 14 A_{[\m_1 \m_2 \m_3 , B}
  A_{\m_4 ] , M} \Theta^N S^B_{N\a} S^{M\a}_A + 8 A_{[\m_1 \m_2 ,
  \a} A_{\m_3 \m_4 ] ,\b} \Theta^M D^\a_{MN} S^{N\b}_A \nonumber \\
  & & \quad - 8 A_{[\m_1
  \m_2 , \a} A_{\m_3 , M} A_{\m_4 ],N} \Theta^P D^\a_{PQ} D^{QM}_\b
  S^{N\b}_A \nonumber \\
  & & \quad
  - \frac{1}{3} A_{[\m_1 , M}A_{\m_2 , N}A_{\m_3 ,
  P}A_{\m_4 ] , Q} \Theta^R D_{\a, R}{}^M D^\a_S{}^N D_\b^{SP}
  S^{Q\b}_A ] \quad .
  \end{eqnarray}
These field strengths transform covariantly under the gauge
transformations
  \begin{eqnarray}
  & & \d A_{\m, M} = a_{\m , M} + \Lambda_N \Theta^N A_{\m , M} - b
  \Lambda_N \Theta^P D_{\a , P}{}^N D^\a_M{}^Q A_{\m ,Q } \nonumber
  \\
  & & \d A_{\m_1 \m_2 , \a} =a_{\m_1 \m_2 , \a} + \frac{1}{2}
  a_{[\m_1 , M} A_{\m_2 ] , N} D^{MN}_\a +2 \Lambda_M \Theta^M
  A_{\m_1 \m_2 , \a} - 8 \Lambda_N \Theta^P D_{\b , P}{}^N
  f^{\b\g}{}_\a A_{\m_1 \m_2 , \g} \nonumber \\
  & & \d A_{\m_1 \m_2 \m_3 , A} = \de_{[\m_1} \Lambda_{\m_2 \m_3 ], A} + a_{[\m_1
  , M} A_{\m_2 \m_3 ] , \a} S^{M\a}_A + \frac{1}{6} a_{[\m_1 , M}
  A_{\m_2 ,N} A_{\m_3 ], P} D_\a^{MN} S^{P\a}_A\nonumber \\
  & & \quad  + 3 \Lambda_M
  \Theta^M A_{\m_1 \m_2 \m_3 , A } - 8 \Lambda_N \Theta^P D_{\a,
  P}{}^N D^\a_A{}^B A_{\m_1 \m_2 \m_3 , B} \quad ,
  \end{eqnarray}
where the parameters $a_{\m_1 ...\m_n ,M_n}$ are as in eq.
(\ref{aintermsoflambda}), which gives
  \begin{eqnarray}
  & & a_{\m , M} = \de_\m \Lambda_M -16 \Theta^N D^\a_{MN}
  \Lambda_{\m , \a} \nonumber \\
  & & a_{\m_1 \m_2 , \a} = \de_{[\m_1} \Lambda_{\m_2 ] , \a} + 14
  \Theta^M S^A_{M\a} \Lambda_{\m_1 \m_2 , A} \quad .
  \end{eqnarray}
This concludes our four-dimensional analysis.

\section{D=5}
The maximal massless supergravity theory in five dimensions was
derived in \cite{sugraD=5}. Its bosonic sector describes 42 scalars
parametrising the manifold $E_{6(+6)}/USp(8)$, the metric and a
1-form in the ${\bf 27}$. This is described by the decomposition of
$E_{11}$ corresponding to the deletion of node 5 in fig. 1. The
level zero generators and the form generators up to the 4-form
included that occur in this decomposition of $E_{11}$ with respect
to $GL(5, \mathbb{R}) \otimes E_6$ are
\cite{fabiopeterE11origin,fabiopeterogievetsky}
  \begin{equation}
  K^\m{}_\n \ ({\bf 1}) \ \ R^\alpha  \ ({\bf 78}) \ \ R^{\m , M} \ ({\bf \overline{27}})
  \ \ R^{\m_1 \m_2}{}_M \ ({\bf 27})
  \ \ R^{\m_1 \m_2 \m_3 , \alpha} \ ({\bf 78}) \
  \ R^{\m_1 \m_2 \m_3 \m_4}{}_{MN} \ ({\bf \overline{351}}) \quad ,
  \label{listofgeninD=5}
  \end{equation}
where we devote in brackets the $E_6$ representation of each
generator.

We use the notations of \cite{hierarchyE11}, which we now partly
review. The commutators between the scalars and the other generators
are dictated by the $E_6$ representation that the generators carry.
The other $E_{11}$ commutation relations which will be relevant in
this paper are \cite{hierarchyE11}
  \begin{eqnarray}
  & & [R^{\m_1 ,M} , R^{\m_2 ,N} ]= d^{MNP} R^{\m_1 \m_2}{}_P \nonumber \\
  & & [R^{\m_1 ,N} , R^{\m_2 \m_3}{}_M ]= g_{\alpha\beta} (D^\alpha )_M{}^N R^{\m_1 \m_2 \m_3, \beta}
  \nonumber \\
  & & [R^{\m_1 \m_2 }{}_M , R^{\m_3 \m_4}{}_N ]= R^{\m_1 \m_2 \m_3 \m_4 }{}_{MN} \nonumber \\
  & & [R^{\m_1 , P}, R^{\m_2 \m_3 \m_4 , \alpha} ]= S^{\alpha P, MN} R^{\m_1 \m_2 \m_3 \m_4}{}_{MN}
  \quad , \label{E11algebrainfivedimensions}
  \end{eqnarray}
where $d^{MNP}$ is the completely symmetric invariant tensor of
$E_6$ and $g^{\alpha\beta}$ is the metric that we use to raise and
lower indices in the adjoint defined by the relation
  \begin{equation}
  D^\alpha_M{}^N D^\beta_N{}^M = g^{\alpha \beta} \quad ,
  \end{equation}
which is proportional to the Cartan-Killing metric as it is evident
from
  \begin{equation}
  f_{\alpha \beta\gamma} f^{\alpha \beta \delta} = - 4
  \delta^\delta_\gamma
  \quad ,
  \end{equation}
where $f^{\alpha \beta \gamma}$ are the structure constants of
$E_6$. The other $E_6$ invariant tensor $ S^{\alpha P, MN}$ that
occurs in eq. (\ref{E11algebrainfivedimensions}) is defined by
  \begin{equation}
  S^{\alpha M , NP} = -3 D^\alpha_Q{}^{[N} d^{P] MQ} \quad
  .\label{S=Ddinfivedimensions}
  \end{equation}
In the following we will also need the conjugate of this invariant
tensor, that is
  \begin{equation}
  S^{\alpha}_{ M , NP} = -3 D^\alpha_{[N}{}^{Q} d_{P] MQ} \quad
  . \label{lowerSE6}
  \end{equation}

We now consider the trombone deformations of this algebra. As
explained in section 2, we consider as our starting point the
commutation relation
  \begin{equation}
  [ R^{\m ,M} , P_\n ] = \delta^\m_\n (\Theta^M K + a \Theta^N
  D_{\a , N}{}^M R^\a )
  \quad ,
  \end{equation}
with the parameter $a$ to be determined. We proceed as in the
previous cases, imposing the closure of the Jacobi identities
involving two form generators and momentum. This determines the
parameter $a$ to be
  \begin{equation}
  a = - \frac{9}{2} \quad ,
  \end{equation}
and it also determines all the commutators of the other form
generators with momentum. Summarising, we get the commutators
  \begin{eqnarray}
  & &   [ R^{\m ,M} , P_\n ] = \delta^\m_\n (\Theta^M K - \frac{9}{2} \Theta^N
  D_{\a , N}{}^M R^\a ) \nonumber \\
  & & [ R^{\m_1 \m_2 }{}_M , P_\n ] = 15 d_{MNP} \Theta^N
  \delta^{[\m_1}_\n R^{\m_2 ] , P} \nonumber \\
  & & [ R^{\m_1 \m_2 \m_3 , \a} , P_\n ] = \frac{27}{2} \Theta^M
  D^\a_M{}^N \delta^{[\m_1 }_\n R^{\m_2 \m_3 ]}{}_N \nonumber \\
  & & [ R^{\m_1 \m_2 \m_3 \m_4}{}_{MN} , P_\n ] = 10 \Theta^P
  S^\a_{P , MN} \delta^{[\m_1}_\n R^{\m_2 \m_3 \m_4 ]}_\a \quad
  , \end{eqnarray}
where the invariant tensor $ S^\a_{P , MN} $ is defined in eq.
(\ref{lowerSE6}).

We now consider the quadratic constraints. The constraint of eq.
(\ref{purespinorconstraint}) becomes in this $E_6$ case
  \begin{equation}
  \Theta^M \Theta^N d_{MNP} = 0 \quad , \label{purespinorE6}
  \end{equation}
which is precisely the $E_6$ ``pure spinor constraint'' obtained in
\cite{henning}. All the other Jacobi identities give constraints
that are all contained in this condition. In particular, the Jacobi
identity involving the scalar generator $\Theta^M K - \frac{9}{2}
\Theta^N
  D_{\a , N}{}^M R^\a$, the 1-form generator and momentum is
implied by eq. (\ref{purespinorE6}) using the $E_6$ relation
  \begin{equation}
  g_{\alpha \beta} D^\alpha_M{}^N D^\beta_P{}^Q = {1 \over 6} \delta^N_P
  \delta_M^Q + {1 \over 18} \delta_M^N \delta_P^Q -{5 \over 3} d^{NQR} d_{MPR}
  \quad . \label{finalDDequationinD=5appendix}
  \end{equation}
whose derivation based on $E_{11}$ is presented in Appendix A of
ref. \cite{hierarchyE11}.

We now determine the field strengths and gauge transformations of
the fields as they result from the deformed local $E_{11}$ group
element. We start from the group element
   \begin{equation}
  g = e^{x \cdot P} e^{A_{\m_1
  ...\m_4}^{MN} R^{\m_1 ... \m_4}_{MN} } e^{A_{\m_1 \m_2 \m_3, \alpha}
  R^{\m_1 \m_2 \m_3, \alpha}} e^{A_{\m_1 \m_2}^M R^{\m_1 \m_2}_M }
  e^{A_{\m
  ,M} R^{\m , M}} e^{\phi_\alpha R^\alpha} e^{h_\m{}^\n K^\m{}_\n} \quad ,
  \label{fivedimgroupelement}
  \end{equation}
and using the general formulae of section 2 as well as the results
in this section we find
  \begin{eqnarray}
  & & F_{\m_1 \m_2 , M} = 2 [ \de_{[\m_1 } A_{\m_2 ],M} -\frac{3}{4}
  \Theta^N A_{[\m_1 , N} A_{\m_2 ] , M} -15 \Theta^N d_{MNP} A_{\m_1
  \m_2 }^P ]\nonumber \\
  & & F_{\m_1 \m_2 \m_3}^M = 3 [ \de_{[\m_1 } A_{\m_2 \m_3 ] }^M +
  \frac{1}{2} \de_{[\m_1} A_{\m_2 , N} A_{\m_3 ] , P} d^{MNP} -
  \frac{27}{2} \Theta^N D^\a_N{}^M A_{\m_1 \m_2 \m_3 , \a} \nonumber \\
  & & \quad - 15
  A_{[\m_1 \m_2 }^N A_{\m_3 ], P} \Theta^Q d_{QNR} d^{RPM} +
  \frac{3}{4} A_{[\m_1 , N} A_{\m_2 , P} A_{\m_3 ], Q} \Theta^R D_{\a
  , R}{}^N D^\a_S{}^P d^{SQM} ] \nonumber \\
  & & F_{\m_1 \m_2 \m_3 \m_4 }^\a = 4 [ \de_{[\m_1 } A_{\m_2 \m_3
  \m_4 ]}^\a - \de_{[\m_1 } A_{\m_2 \m_3}^M A_{\m_4 ] , N}
  D^\a_M{}^N - \frac{1}{6} \de_{[\m_1} A_{\m_2 , M } A_{\m_3 , N}
  A_{\m_4 ], P} d^{MNQ} D^\a_Q{}^P \nonumber \\
  & & \quad -10 \Theta^P S^\a_{P,MN} A_{\m_1
  ...\m_4 }^{MN} + \frac{27}{2} A_{[\m_1 \m_2 \m_3 , \b} A_{\m_4 ] ,
  M} \Theta^N D^\b_N{}^P D^\a_P{}^M \nonumber \\
  & & \quad - \frac{15}{2} A_{[\m_1 \m_2 }^M
  A_{\m_3 \m_4 ]}^N \Theta^P d_{PMQ} D^\a_N{}^Q + \frac{15}{2}
  A_{[\m_1 \m_2}^M A_{\m_3 , N} A_{\m_4 ] , P }\Theta^Q d_{QMR}
  d^{RNS} D^\a_S{}^P \nonumber \\
  & & \quad - \frac{3}{16} A_{[\m_1 , M} A_{\m_2 , N}
  A_{\m_3 , P } A_{\m_4 ] , Q} \Theta^R D_{\b , R}{}^M D^\b_S{}^N
  d^{SPT} D^\a_T{}^Q ] \quad .
  \end{eqnarray}
These field strengths are covariant with respect to the gauge
transformations
  \begin{eqnarray}
  & & \delta A_{\m , M} = a_{\m , M} + \Lambda_N \Theta^N A_{\m , M}
  - \frac{9}{2} \Lambda_N \Theta^P D_{\a , P}{}^N D^\a_M{}^Q A_{\m ,
  Q} \nonumber \\
  & & \delta A_{\m_1 \m_2}^M = a_{\m_1 \m_2}^M + \frac{1}{2}
  a_{[\m_1 , N} A_{\m_2 ] , P} d^{MNP} + 2 \Lambda_N \Theta^N
  A_{\m_1 \m_2 }^M + \frac{9}{2} \Lambda_N \Theta^P D_{\a , P}{}^N
  D^\a_Q{}^M A_{\m_1 \m_2 }^Q \nonumber \\
  & & \d A_{\m_1 \m_2 \m_3 }^\a = \de_{[\m_1 } A_{\m_2 \m_3 ]}^\a +
  a_{[\m_1 , M} A_{\m_2 \m_3 ]}^N D^\a_N{}^M + \frac{1}{6} a_{[\m_1
  , M} A_{\m_2 , N} A_{\m_3 ] , P } d^{MNQ} D^\a_Q{}^P \nonumber \\
  & & \quad + 3 \Lambda_M
  \Theta^M A_{\m_1 \m_2 \m_3 }^\a - \frac{9}{2} \Lambda_N \Theta^P
  D_{\b , P}{}^N f^{\b\g}{}_\a A_{\m_1 \m_2 \m_3 , \g} \quad ,
  \end{eqnarray}
where using eq. (\ref{aintermsoflambda}) one obtains the parameters
$a_{\m_1 ...\m_n ,M_n}$ in terms of the gauge parameters as
  \begin{eqnarray}
  & & a_{\m , M}  = \de_{\m} \Lambda_M + 15 \Theta^N d_{MNP }
  \Lambda_\m^P \nonumber \\
  & & a_{\m_1 \m_2}^M = \de_{[\m_1} \Lambda_{\m_2 ]}^M + \frac{27}{2}
  \Theta^N D^\a_N{}^M \Lambda_{\m_1 \m_2 , \a} \quad .
  \end{eqnarray}

We now consider the six-dimensional case.

\section{D=6}
The symmetry of the massless maximal supergravity theory in 6
dimensions \cite{taniisixdim} is $SO(5,5)$, and the bosonic sector
of the theory describes 25 scalars parametrising the symmetric
manifold $SO(5,5)/[SO(5) \times SO(5)]$, the metric, a 1-form in the
${\bf 16}$ and a 2-form in the ${\bf 10}$, whose field strength
satisfies a self-duality condition. From $E_{11}$ this theory arises
after deleting node 6 in the $E_{11}$ Dynkin diagram of fig. 1.

In this paper we will only consider the form generators of rank up
to four included. For a more detailed analysis including the 5-form
generators we refer to \cite{hierarchyE11}, whose notation we fully
adopt. Using the results of \cite{hierarchyE11}, the analysis
presented here can easily be extended to higher rank form
generators. The level zero generators and the form generators up to
the 4-form included are
  \begin{equation}
  K^\m{}_\n \ ({\bf 1}) \ \  R^{MN} \ ({\bf {45}}) \  \ R^{\m , \dot{\alpha}} \ ({\bf
  \overline{16}} )
  \ \  R^{\m_1 \m_2, M} \ ( {\bf 10} ) \ \  R^{\m_1 \m_2 \m_3 , {\alpha}}
  \  ({\bf 16} )\ \  R^{\m_1 \m_2 \m_3 \m_4, MN} \  ({\bf 45} )
  \quad , \label{genlistD=6}
  \end{equation}
where we denote in brackets the corresponding $SO(5,5)$
representations.

We review now the $SO(5,5)$ Gamma matrix conventions of
\cite{hierarchyE11}. We are using a Weyl basis, so that the Gamma
matrices have the form
  \begin{equation}
  \Gamma_{M ,A}{}^B =  \left( \begin{array}{cc}
  0  &  \Gamma_{M, \alpha}{}^{\dot{\beta}} \\
  \Gamma_{M, \dot{\alpha}}{}^\beta & 0 \end{array} \right) \quad ,
  \end{equation}
where $A=1, ..., 32$. They satisfy the Clifford algebra
  \begin{equation}
  \{ \Gamma_M , \Gamma_N \} = 2 \eta_{MN}
  \end{equation}
where $\eta_{MN}$ is the $SO(5,5)$ Minkowski metric. The charge
conjugation matrix is
  \begin{equation}
  C^{AB} =  \left( \begin{array}{cc}
  0 &   C^{\alpha \dot{\beta}}  \\
  C^{\dot{\alpha} \beta} & 0 \end{array} \right) \quad ,
  \label{thematrixC}
  \end{equation}
which is antisymmetric and unitary, that is
  \begin{equation}
  C^{\alpha \dot{\beta}} =-  C^{\dot{\beta} \alpha}
  \end{equation}
and
  \begin{equation}
  C^\dagger_{\alpha \dot{\beta}} C^{\dot{\beta} \gamma} =
  \delta_\alpha^\gamma \qquad   C^\dagger_{\dot{\alpha}{\beta}}
  C^{{\beta} \dot{\gamma}} = \delta_{\dot{\alpha}}^{\dot{\gamma}}
  \quad ,
  \end{equation}
and satisfies the property
  \begin{equation}
  C \Gamma_M C^\dagger = - \Gamma_M^T \quad .
  \end{equation}

We now review the commutation relations of \cite{hierarchyE11}. The
$SO(5,5)$ algebra is
  \begin{equation}
  [ R^{MN} , R^{PQ} ] = \eta^{MP} R^{NQ} - \eta^{NP} R^{MQ} +
  \eta^{NQ} R^{MP} - \eta^{MQ} R^{NP}
  \end{equation}
while the commutators of the $SO(5,5)$ generators with the positive
level generators are
  \begin{eqnarray}
  & & [ R^{MN} , R^{\m ,\dot{\alpha}} ] = - {1 \over 2}
  \Gamma^{MN}{}_{\dot{\beta}}{}^{\dot{\alpha}} R^{\m , \dot{\beta}}
  \nonumber \\
  & & [ R^{MN} , R^{\m_1 \m_2 ,P} ] = \eta^{MP} R^{\m_1 \m_2 , N} - \eta^{NP} R^{\m_1 \m_2 ,
  M}
  \nonumber \\
  & &
  [ R^{MN} , R^{\m_1 \m_2 \m_3 ,\alpha} ] = - {1 \over 2}
  \Gamma^{MN}{}_{\beta}{}^{\alpha} R^{\m_1 \m_2 \m_3 , {\beta}} \quad .
  \end{eqnarray}
The commutators of the positive level generators of eq.
(\ref{genlistD=6}) are
  \begin{eqnarray}
  & & [ R^{ \m_1 ,\dot{\alpha}} , R^{\m_2 , \dot{\beta}} ] = ( C \Gamma_M )^{
  \dot{\alpha} \dot{\beta}} R^{\m_1 \m_2 , M}
  \nonumber \\
  & & [ R^{ \m_1 ,\dot{\alpha}} , R^{\m_2 \m_3 , M} ] =
  \Gamma^M{}_\alpha{}^{\dot{\alpha}}
  R^{\m_1 \m_2 \m_3 , \alpha}
  \nonumber \\
  & &
  [ R^{ \m_1 \m_2 , M} , R^{\m_3 \m_4 , N} ] = R^{\m_1 ... \m_4  , MN}
  \nonumber \\
  & & [ R^{ \m_1 ,\dot{\alpha}} , R^{\m_2 \m_3 \m_4 , \alpha} ] = {1 \over 4}( C \Gamma_{MN} )^{
  \dot{\alpha} \alpha} R^{\m_1 ... \m_4  , MN}
  \quad .
  \end{eqnarray}

We now consider the trombone deformations of the $E_{11,6}^{local}$
algebra. We start from the commutators of eq. (\ref{defalgebra1})
and (\ref{defalgebra2}). Using $SO(5,5)$ Fierz identities, one can
show that the Jacobi identities involving two positive level
generators and the momentum generator give
  \begin{eqnarray}
  & & [ R^{\m , \dot{\a} } , P_\n ] = \delta^\m_\n [ \Theta_\a C^{\a
  \dot{\a}} K - \frac{1}{5} \Theta_\a ( C \Gamma^{MN} )^{\a
  \dot{\a}} R_{MN} ] \nonumber \\
  & & [ R^{\m_1 \m_2 , M} , P_\n ] = - \frac{4}{5} \Theta_\a
  \Gamma^M_{\dot{\a}}{}^\a \delta^{[\m_1}_\n R^{\m_2 ] , \dot{\a}}
  \nonumber \\
  & & [ R^{\m_1 \m_2 \m_3 , \a} , P_\n ] = \frac{6}{5} \Theta_\beta
  ( C \Gamma_M )^{\a \b} \delta^{[\m_1}_\n R^{\m_2 \m_3 ] , M}
  \nonumber \\
  & & [ R^{\m_1 ...\m_4 , MN} , P_\n ] = \frac{8}{5} \Theta_\a (
  \Gamma^{MN} )_\b{}^\a \delta^{[\m_1}_\n R^{\m_2 \m_3 \m_4 ] ,
  \beta } \quad ,
  \end{eqnarray}
where the embedding tensor $\Theta_\a$ is in the same representation
of the 1-form generator $R^{\m , \dot{\a}}$ because spinor indices
are raised and lowered using the matrix $C$ of eq.
(\ref{thematrixC}).

We then consider the quadratic constraints. Eq.
(\ref{purespinorconstraint}), applied to this case, gives
  \begin{equation}
  \Theta_\a \Theta_\b ( C \Gamma_M )^{\a\b} = 0 \quad ,
  \end{equation}
which is precisely the $SO(5,5)$ pure spinor constraint of
\cite{henning}. It can be shown that all the other Jacobi identities
close automatically using this equation.

To conclude this section, we determine the field strengths and gauge
transformations of the fields as they result from applying the
formulae of section 2 to this six-dimensional case. We get the field
strengths
  \begin{eqnarray}
  & & F_{\m_1 \m_2 , \dot{\a}} = 2 [ \de_{[\m_1} A_{\m_2 ] ,
  \dot{\a}}- \frac{1}{2} \Theta_\b C^{\b \dot{\b}} A_{[\m_1 ,
  \dot{\b} }A_{\m_2 ] , \dot{\a}} - \frac{1}{20} A_{[\m_1 ,
  \dot{\g}} A_{\m_2 ] , \dot{\d}}\Theta_\b ( C \Gamma^{MN} )^{\b
  \dot{\g}} \Gamma_{MN , \dot{\a}}{}^{\dot{\d}} \nonumber \\
  & &  \ + \frac{4}{5} \Theta_\a \Gamma^M_{\dot{\a}}{}^\a A_{\m_1
  \m_2 ,M} ]\nonumber \\
  & & F_{\m_1 \m_2 \m_3 , M} = 3 [ \de_{[\m_1} A_{\m_2 \m_3 ], M} +
  \frac{1}{2} \de_{[\m_1} A_{\m_2 , \dot{\a}} A_{\m_3 ] , \dot{\b}}
  (C \Gamma_M )^{\dot{\a} \dot{\b}} - \frac{6}{5} \Theta_\b (C
  \Gamma_M )^{\a\b} A_{\m_1 \m_2 \m_3 , \a} \nonumber \\
  & & \ + \frac{4}{5} A_{[\m_1
  \m_2 ,N} A_{\m_3 ] , \dot{\a}}\Theta_\b \Gamma^N_{\dot{\b}}{}^\b (
  C \Gamma_M )^{\dot{\a} \dot{\b}} - \frac{1}{60} A_{[\m_1,
  \dot{\a}} A_{\m_2 , \dot{\b}} A_{\m_3 ] , \dot{\g}} \Theta_\a (C
  \Gamma^{NP} )^{\a\dot{\a}} (\Gamma_{NP})_{\dot{\d}}{}^{\dot{\b}} (
  C \Gamma_M )^{\dot{\d} \dot{\g}} ]\nonumber \\
  & & F_{\m_1 \m_2 \m_3  \m_4 , \a} = 4 [ \de_{[\m_1 } A_{\m_2 \m_3
  \m_4 ] ,\a} - \de_{[\m_1} A_{\m_2 \m_3 , M} A_{\m_4 ] , \dot{\a}}
  \Gamma^M_\a{}^{\dot{\a}} \nonumber \\
  & & \ - \frac{1}{6} \de_{[\m_1} A_{\m_2 ,
  \dot{\a} } A_{\m_3 , \dot{\b} } A_{\m_4 ] ,\dot{\g}} (C \Gamma_M
  )^{\dot{\a}\dot{\b}} \Gamma^M_\a{}^{\dot{\g} } - \frac{8}{5}
  \Theta_\b \Gamma^{MN}_\a{}^\b A_{\m_1 ...\m_4 , MN} \nonumber \\
  & & \ + \frac{6}{5}
  A_{[\m_1 \m_2 \m_3 , \b} A_{\m_4 ] , \dot{\a}} \Theta_\g (C
  \Gamma_M )^{\b\g} \Gamma^M_\a{}^{\dot{\a}} + \frac{2}{5} A_{[\m_1
  \m_2 , M} A_{\m_3 \m_4 ] ,N} \Theta_\b \Gamma^M_{\dot{\a}}{}^\b
  \Gamma^N_\a{}^{\dot{\a}} \nonumber \\
  & & \ -\frac{2}{5} A_{[\m_1 \m_2 , M} A_{\m_3 ,
  \dot{\a}}A_{\m_4 ] , \dot{\b}} \Theta_\b \Gamma^M_{\dot{\g}}{}^\b
  (C \Gamma_N )^{\dot{\g}\dot{\a}} \Gamma^N_\a{}^{\dot{\b}}
  \nonumber \\
  & & \ + \frac{1}{240} A_{[\m_1 , \dot{\a}} A_{\m_2 , \dot{\b}}
  A_{\m_3 , \dot{\g}}A_{\m_4 ] ,\dot{\d}}\Theta_\b  ( C \Gamma^{MN}
  )^{\b\dot{\a}} \Gamma_{MN, \dot{\e}}{}^{\dot{\b}} (C \Gamma_P
  )^{\dot{\e}\dot{\g}} \Gamma^P_\a{}^{\dot{\d}}]
  \end{eqnarray}
and the gauge transformations
    \begin{eqnarray}
  & & \delta A_{\m, \dot{\alpha}} = a_{\m , \dot{\alpha}} - {1 \over 2}
  a^{MN} \Gamma_{MN, \dot{\alpha}}{}^{\dot{\beta}} A_{\m ,
  \dot{\beta}} + \Theta_\a C^{\a \dot{\b}} \Lambda_{\dot{\b}} A_{\m ,\dot{\a}}
  \nonumber \\
  & & \delta A_{\m_1 \m_2 ,M} = a_{\m_1 \m_2 , M} - {1 \over 2} ( C \Gamma_M
  )^{\dot{\alpha} \dot{\beta} } A_{[\m_1 ,\dot{\alpha} } a_{\m_2 ] ,
  \dot{\beta}} + 2 a_M{}^N A_{\m_1 \m_2 ,N}  + 2 \Theta_\a \Lambda_{\dot{\a}}C^{\a \dot{\a}} A_{\m_1 \m_2 ,
  M}\nonumber \\
   & & \delta A_{\m_1 \m_2 \m_3 , \alpha} = \de_{[\m_1} \Lambda_{ \m_2 \m_3 ], \alpha} +
   \Gamma^M_\alpha{}^{\dot{\alpha}} A_{[\m_1 \m_2 , M} a_{\m_3 ],
   \dot{\alpha}} - { 1 \over 3!} (C \Gamma_M )^{\dot{\beta}
   \dot{\gamma}} \Gamma^M_\alpha{}^{\dot{\alpha}} A_{[\m_1 ,
   \dot{\alpha}} A_{\m_2 , \dot{\beta}} a_{\m_3 ] , \dot{\gamma}} \nonumber \\
   & & \quad \qquad  - {1
   \over 2 } a^{MN} \Gamma_{MN, \alpha}{}^\beta A_{\m_1 \m_2 \m_3 ,
   \beta} + 3 \Theta_{\b} C^{\b \dot{\a}} \Lambda_{\dot{\a}} A_{\m_1 \m_2 \m_3 , \a}\quad ,
   \end{eqnarray}
where using eq. (\ref{aintermsoflambda}) we express the parameters
$a_{\m_1 ...\m_n ,M_n}$ in terms of the gauge parameters as
  \begin{eqnarray}
  & & a_{MN} = - \frac{1}{5} \Lambda_{\dot{\a}}\Theta_\a ( C
  \Gamma_{MN} )^{\a \dot{\a}} \nonumber \\
  & & a_{\m , \dot{\a}} = \de_{\m} \Lambda_{\dot{\a}} - \frac{4}{5}
  \Theta_\a \Gamma^M_{\dot{\a}}{}^\a \Lambda_{\m , M} \nonumber \\
  & & a_{\m_1 \m_2 ,M} = \de_{[\m_1} \Lambda_{\m_2 ] ,M} +
  \frac{6}{5} \Theta_\b (C \Gamma_M )^{\a\b} \Lambda_{\m_1 \m_2 ,
  \a} \quad .
  \end{eqnarray}
This concludes the six-dimensional analysis.

\section{D=7}
The massless maximal supergravity theory in 7 dimensions
\cite{maxsugrasevendim} has a bosonic sector containing 14 scalars
parametrising $SL(5,\mathbb{R})/SO(5)$, the metric, a 1-form in the
${\bf \overline{10}}$ and a 2-form in the ${\bf 5}$ of
$SL(5,\mathbb{R})$. This theory results from $E_{11}$ after deletion
of node 7 in fig. 1. As in all other cases, we consider the level
zero generators as well as the positive level generators up to rank
four included. These are
  \begin{equation}
  K^\m{}_\n \ ({\bf 1}) \  \ R^M{}_N  \ ({\bf 24}) \ \  R^{\m , MN} \ ({\bf 10}) \ \  R^{\m_1 \m_2}{}_M \ ({\bf \overline{5}})
  \ \  R^{\m_1 \m_2 \m_3 , M} \  ({\bf 5}) \  \
  R^{\m_1 \m_2 \m_3 \m_4}{}_{MN} \  ({\bf \overline{10}}) \label{poslevgenD=7}
  \quad ,
  \end{equation}
where we denote in brackets the corresponding $SL(5 , \mathbb{R})$
representation.

We now review the algebraic relations of \cite{hierarchyE11} between
the generators in eq. (\ref{poslevgenD=7}). The relevant commutators
are those involving the generators of $SL(5, \mathbb{R})$,
  \begin{eqnarray}
  & &   [ R^M{}_N , R^P{}_Q ] = \delta^P_N R^M{}_Q - \delta^M_Q
  R^P{}_N \nonumber \\
  & & [ R^M{}_N , R^{\m , PQ} ]= \delta^P_N R^{\m , MQ} + \delta^Q_N
  R^{\m, PM} - {2 \over 5} \delta^M_N R^{\m, PQ}
  \nonumber \\
  & & [ R^M{}_N , R^{\m_1 \m_2}{}_P ] = - \delta^M_P R^{\m_1 \m_2}{}_N + {1 \over 5}
  \delta^M_N R^{\m_1 \m_2}{}_P
  \end{eqnarray}
as well as those between the positive level generators
  \begin{eqnarray}
  & & [ R^{\m_1 , MN} , R^{\m_2 , PQ} ] = \epsilon^{MNPQR} R^{\m_1
  \m_2}{ }_R
  \nonumber \\
  & & [ R^{\m_1 , MN} , R^{\m_2 \m_3}{}_{ P} ] = \delta^{[M}_P R^{\m_1 \m_2
  \m_3 , N]}
  \nonumber \\
  & & [ R^{\m_1 \m_2}{}_M , R^{\m_3 \m_4}{}_N ] = R^{\m_1 ...\m_4 }{}_{MN}
  \nonumber \\
  & &
  [ R^{\m_1 , MN} , R^{\m_2 \m_3 \m_4, P}] = \epsilon^{MNPQR} R^{\m_1
  ...\m_4}{}_{QR}
 \quad .
  \end{eqnarray}
To prove that all the Jacobi identities close one makes use of
  \begin{equation}
  \epsilon^{M_1 ..M_5} \epsilon_{N_1 ...N_5} = 5! \delta^{[M_1
  ...M_5]}_{[N_1 ...N_5 ]} \quad .
  \end{equation}

We now consider the trombone deformations of the $E_{11,7}^{local}$
algebra as they result from applying to this $SL(5 ,\mathbb{R})$
case the generic commutators of eq. (\ref{defalgebra1}) and
(\ref{defalgebra2}). Imposing the closure of the Jacobi identities
involving two positive level generators and momentum one obtains
   \begin{eqnarray}
   & & [ R^{\m , MN} , P_\n ] = \delta^\m_\n [ \Theta^{MN} K -
   \frac{5}{3} \Theta^{[M | P | } R^{N ]}{}_P ] \nonumber \\
   & & [ R^{\m_1 \m_2 }{}_M , P_\n ] = \frac{5}{12} \e_{MNPQR}
   \Theta^{NP} \delta^{[\m_1}_\n R^{\m_2 ] , QR} \nonumber \\
   & & [ R^{\m_1 \m_2 \m_3 , M} , P_\n ] = - 5 \Theta^{MN}
   \delta^{[\m_1}_\n R^{\m_2 \m_3 ]}{}_N \nonumber \\
   & & [ R^{\m_1 \m_2 \m_3 \m_4}{}_{MN} , P_\n ] = \frac{5}{6}
   \e_{MNPQR} \Theta^{PQ} \delta^{[\m_1}_\n R^{\m_2 \m_3 \m_4 ] , R}
   \quad .
   \end{eqnarray}
The quadratic constraint of eq. (\ref{purespinorconstraint}), when
applied to this case, gives
  \begin{equation}
  \Theta^{MN} \Theta^{PQ} \e_{MNPQR} = 0 \quad ,
  \end{equation}
and one can show that this condition in enough to guarantee the
closure of all the Jacobi identities.

To conclude this section, we now determine the field strengths and
gauge transformations of the fields. Starting from the group element
of the form of eq. (\ref{generalgroupelement}) and using eqs.
(\ref{generalformoffieldstrengths}) and
(\ref{generalformofgaugetransfs}) we get the field strength of the
vectors
  \begin{equation}
  F_{\m_1 \m_2, MN} = 2[\partial_{[\m_1} A_{\m_2 ] , MN}- \frac{5}{12}
  \Theta^{PQ} \e_{MNPQR } A_{\m_1 \m_2 }^R + \frac{1}{2} \Theta^{PQ} A_{[\m_1 , PQ} A_{\m_2 ] , MN}  ] \quad ,
  \end{equation}
the field strength of the 2-form
  \begin{eqnarray}
  & & F_{\m_1 \m_2 \m_3 }^M = 3 [ \partial_{[\m_1} A_{\m_2 \m_3]}^M + {1
  \over 2} \epsilon^{MNPQR} \partial_{[\m_1} A_{\m_2 , NP } A_{\m_3 ] ,
  QR} -5  \Theta^{MN} A_{\m_1 \m_2 \m_3 , N} \nonumber \\
  & & \quad\qquad - {5 \over 12} \Theta^{RS}
  \epsilon^{MTUPQ} \e_{NRSTU} A_{[\m_1 \m_2}^N A_{\m_3 ] , PQ} ]
  \end{eqnarray}
and the field strength of the 3-form
  \begin{eqnarray}
  & & F_{\m_1 ...\m_4 , M} = 4[ \partial_{[\m_1} A_{\m_2 ...\m_4 ], M} -
  \partial_{[\m_1} A_{\m_2 \m_3}^N A_{\m_4 ] ,NM} -{1 \over 6}
  \partial_{[\m_1} A_{\m_2, NP} A_{\m_3 , QR} A_{\m_4 ] , SM}
  \epsilon^{SNPQR} \nonumber \\
  & & \quad \qquad - \frac{5}{6} A_{\m_1 ...\m_4}^{NP} \Theta^{QR} \e_{MNPQR}
  - 5 A_{[\m_1 \m_2 \m_3 , P} A_{\m_4 ] ,QM} \Theta^{PQ} \nonumber \\
  & & \quad \qquad + \frac{5}{24} A_{[\m_1 \m_2}^R A_{\m_3 , NP} A_{\m_4 ] , QM} \Theta^{ST} \e_{RSTUV} \e^{UVNPQ}] \quad ,
  \end{eqnarray}
that are covariant under the gauge transformations
  \begin{eqnarray}
  & & \delta A_{\m, MN} = a_{\m, MN} + 2 a_{[M}{}^P A_{\m, \vert P
  \vert N]} + \Lambda_{PQ} \Theta^{PQ} A_{\m , MN} \nonumber \\
  & & \delta A_{\m_1 \m_2}^M = a_{\m_1 \m_2}^M - {1 \over 2}
  \epsilon^{MNPQR} A_{[\m_1, NP} a_{\m_2 ], QR} - a_N{}^M A_{\m_1
  \m_2}^N  + 2 \Lambda_{NP} \Theta^{NP} A_{\m_1 \m_2}^M \nonumber \\
  & & \delta A_{\m_1 \m_2 \m_3 , M} = \de_{[\m_1} \Lambda_{\m_2 \m_3 ] ,M} + A_{[\m_1
  \m_2}^N a_{\m_3], NM} +{1 \over 3!} \epsilon^{NQRST} A_{[\m_1, MN}
  A_{\m_2, QR} a_{\m_3 ] , ST} \nonumber \\
  & & \quad \qquad + a_M{}^N A_{\m_1 \m_2 \m_3 , N} + 3 \Lambda_{NP}
  \Theta^{NP} A_{\m_1 \m_2 \m_3 , M} \quad ,
  \end{eqnarray}
where the parameters $a$ are given in terms of the gauge parameters
as
  \begin{eqnarray}
  & & a_M{}^N =  \frac{5}{3} \Lambda_{MP} \Theta^{PN}  \nonumber \\
  & & a_{\m, MN} =\partial_{\m} \Lambda_{MN} + \frac{5}{12} \Theta^{PQ} \e_{MNPQR} \Lambda_{\m}^R \nonumber \\
  & & a_{\m_1 \m_2}^M = \partial_{[ \m_1} \Lambda_{\m_2 ]}^M +  5 \Theta^{MN} \Lambda_{\m_1 \m_2,
  N} \quad .
  \end{eqnarray}

\section{D=8}
The maximal massless eight-dimensional supergravity was derived in
\cite{salamsezginD=8}. Its bosonic sector contains seven scalars
parametrising the manifold $SL(3,\mathbb{R})/SO(3) \times
SL(2,\mathbb{R})/SO(2)$, the metric, a vector in the ${\bf
(\overline{3},2)}$ of the internal symmetry group $SL(3,\mathbb{R})
\times SL(2,\mathbb{R})$, a 2-form in ${\bf (3,1)}$ and an
$SL(2,\mathbb{R})$ doublet of 3-forms which satisfy self-duality
conditions. Theory corresponds to the decomposition of $E_{11}$
which results from deleting node 8 in the Dynkin diagram of fig. 1.

As in all other cases, we consider the level zero generators and the
positive level form generators with up to four indices included.
These are
    \begin{eqnarray}
  & & R^\m{}_\n \ \ {\bf (1,1)} \quad  R^i  \ \ {\bf (1,3)} \quad   R^M{}_N \  \ {\bf (8,1)}
  \quad   R^{\m , M \alpha}  \ \ {\bf (3,2)} \nonumber \\
  & & R^{\m_1 \m_2}{}_M   \ \ {\bf (\overline{3},1)} \quad  R^{\m_1 \m_2
\m_3 , \alpha}
  \  \ {\bf (1,2)} \quad  R^{\m_1 \m_2 \m_3 \m_4, M}  \ \ {\bf (3,1)}  \quad , \label{eightdimlistofgen}
  \end{eqnarray}
where we have denoted in brackets their corresponding
$SL(3,\mathbb{R}) \times SL(2,\mathbb{R})$ representation.

We now review the undeformed algebra of \cite{hierarchyE11}. The
commutators involving the scalar generators are
  \begin{eqnarray}
  & & [R^i , R^j ] = f^{ij}{}_k R^k
  \nonumber \\
  & & [ R^M{}_N , R^P{}_Q ] = \delta^P_N R^M{}_Q - \delta^M_Q
  R^P{}_N \nonumber \\
  & & [ R^i , R^{\m, M\alpha} ] = D^i_\beta{}^\alpha R^{\m, M \beta}
  \nonumber \\
  & & [ R^M{}_N , R^{\m , P\alpha} ] = \delta^P_N R^{\m , M \alpha} - {1
  \over 3} \delta^M_N R^{\m , P \alpha}
  \end{eqnarray}
and similarly for the higher rank form generators. Here
$D^i_\beta{}^\alpha$ are the generators of $SL(2, \mathbb{R})$
satisfying
  \begin{equation}
  [ D^i , D^j ]_\beta{}^\alpha = f^{ij}{}_k D^k_\beta{}^\alpha
  \end{equation}
and $f^{ij}{}_k$ are the structure constants of $SL(2,\mathbb{R})$.
We raise and lower $SL(2,\mathbb{R})$ indices using the
antisymmetric metric $\epsilon^{\alpha\beta}$, that is, for a
generic doublet $V^\alpha$,
  \begin{equation}
  V^\alpha = \epsilon^{\alpha \beta} V_\beta \qquad \quad V_\alpha =
  V^\beta \epsilon_{\beta \alpha} \quad , \label{sl2raiseandlower}
  \quad .
  \end{equation}
The generators
  \begin{equation}
  D^{i, \alpha \beta} = \epsilon^{\alpha \gamma} D^i_\gamma{}^\beta
  \end{equation}
are symmetric in $\alpha\beta$. Useful identities relating the
$SL(2, \mathbb{R})$ generators are
  \begin{equation}
  D_i^{\alpha \beta} D^{i , \gamma \delta }= - {1 \over 4} [
  \epsilon^{\alpha \gamma} \epsilon^{\beta \delta} +
  \epsilon^{\alpha \delta } \epsilon^{\beta \gamma} ]
  \end{equation}
and
  \begin{equation}
  D^i_\beta{}^\gamma D^j_\gamma{}^\alpha + D^j_\beta{}^\gamma
  D^i_\gamma{}^\alpha = {1 \over 2} g^{ij} \delta^\alpha_\beta \quad
  ,\label{symmetricproductsl2}
  \end{equation}
where $g^{ij}$ is the $SL(2, \mathbb{R})$ Killing metric. The
commutators involving the non-scalar generators of eq.
(\ref{eightdimlistofgen}) are
  \begin{eqnarray}
  & & [ R^{\m_1 , M\alpha} , R^{\m_2, N \beta} ] = \epsilon^{\alpha \beta}
  \epsilon^{MNP} R^{\m_1 \m_2}{}_P
  \nonumber \\
  & & [ R^{\m_1 , M\alpha} , R^{\m_2 \m_3 }{}_N ]= \delta^M_N R^{\m_1 \m_2 \m_3
  , \alpha}
  \nonumber \\ & &
  [ R^{\m_1 \m_2}{}_M , R^{\m_3 \m_4 }{}_N ]= \epsilon_{MNP} R^{\m_1 \m_2
  \m_3 \m_4, P}
  \nonumber \\ & &
  [ R^{\m_1 , M\alpha} , R^{\m_2 \m_3 \m_4 , \beta} ]  = -\epsilon^{\alpha \beta} R^{\m_1 \m_2 \m_3
  \m_4 , M}
 \quad .
  \label{masslessE11algebraD=8}
  \end{eqnarray}
One can show that all Jacobi identities are satisfied. This requires
the use of the identities of eqs.
(\ref{sl2raiseandlower})-(\ref{symmetricproductsl2}), as well as the
identities
  \begin{equation}
  \epsilon^{M_1 M_2 M_3} \epsilon_{N_1 N_2 N_3} = 6 \delta^{[M_1
  M_2 M_3]}_{[N_1 N_2 N_3]}
  \end{equation}
and
  \begin{equation}
  \epsilon^{\alpha \beta} \epsilon_{\gamma \delta }V^\gamma W^\delta
  = V^\alpha W^\beta - V^\beta W^\alpha \quad ,
  \end{equation}
where in the last equation $V$ and $W$ are two generic $SL(2,
\mathbb{R})$ doublets.

Following the general analysis of section 2, we now consider the
deformations of the $E_{11,8}^{local}$ algebra as they result from
applying to this case the generic commutators of eq.
(\ref{defalgebra1}) and (\ref{defalgebra2}). Imposing the closure of
the Jacobi identities involving two positive level generators and
momentum one obtains
  \begin{eqnarray}
  & & [ R^{\m , M\a} , P_\n ] = \delta^\m_\n [ \Theta^{M\a} K -12
  \Theta^{M\b} D_{i , \b}{}^\a R^i + 3 \Theta^{N\a} R^M{}_N ]
  \nonumber \\
  & & [ R^{\m_1 \m_2 }{}_M , P_\n ] = 3 \e_{MNP} \e_{\a\b}
  \Theta^{N\a} \delta^{[\m_1}_\n R^{\m_2 ] , P \b} \nonumber \\
  & & [ R^{\m_1 \m_2 \m_3 , \a} , P_\n ] = 0 \nonumber \\
  & & [ R^{\m_1 \m_2 \m_3 \m_ 4, M} , P_\n ] = - 6 \e_{\a\b}
  \Theta^{M\a} \delta^{[\m_1 }_\n R^{\m_2 \m_3 \m_4 ] , \b} \quad ,
  \end{eqnarray}
while all other Jacobi identities close provided that the
eight-dimensional case of the quadratic constraint of eq.
(\ref{purespinorconstraint}), which is
  \begin{equation}
  \Theta^{M\a} \Theta^{N\b} \e_{MNP} \e_{\a\b} = 0 \quad ,
  \end{equation}
holds.

Finally, we determine the field strengths and gauge transformations
of the fields. Starting from the group element of eq.
(\ref{generalgroupelement}) and using eqs.
(\ref{generalformoffieldstrengths}) and
(\ref{generalformofgaugetransfs}) we determine the field strength of
the vectors
  \begin{eqnarray}
  & & F_{\m_1 \m_2 , M \alpha} = 2[ \partial_{[ \m_1} A_{\m_2 ] ,
  M\alpha} - 3 \Theta^{N \b} \e_{\a\b} \e_{MNP} A_{\m_1 \m_2 }^P
  -\frac{3}{2} \e^{\b\g} A_{[\m_1 , M\b} A_{\m_2 ] , N\g}
  \Theta^N_\a \nonumber \\
  & & \quad - 3 A_{[\m_1 , M\b} A_{\m_2 ] , N\a} \Theta^{N\b} ]\quad ,
  \end{eqnarray}
the field strength of the 2-form
  \begin{eqnarray}
  & & F_{\m_1 \m_2 \m_3}^M = 3[ \partial_{[\m_1}A_{\m_2 \m_3]}^M +{1 \over 2 }
  \epsilon^{\alpha \beta} \epsilon^{MNP} A_{[\m_1 ,N\alpha}
  \partial_{\m_2}A_{\m_3 ] , P\beta} -3 \Theta^{N\alpha} A_{[\m_1
  \m_2}^R A_{\m_3 ] , Q\a} \e_{RNP} \e^{PQM}\nonumber \\
  & & - \frac{1}{2} A_{[\m_1 , N \b} A_{\m_2 , P \g} A_{\m_3 ] , Q
  \a} \e^{\a\b} \e^{MPQ} \Theta^{N \g} ]
  \end{eqnarray}
and the field strength of the 3-form
  \begin{eqnarray}
  & & F_{\m_1 ...\m_4 , \alpha} = 4 [ \partial_{[\m_1}A_{\m_2 \m_3 \m_4 ]
  , \alpha} +A_{[\m_1 , M\alpha} \partial_{\m_2} A_{\m_3 \m_4 ]}^M +{1
  \over 3!} \epsilon^{\beta\gamma} \epsilon^{MNP} A_{[\m_1 , M\alpha}
  A_{\m_2 , N\beta} \partial_{\m_3} A_{\m_4 ] , P \gamma}\nonumber \\
  & & \quad - 6 A_{\m_1 ...\m_4 , M} \Theta^{M\b} \e_{\a\b} -
  \frac{3}{2} A_{[\m_1 \m_2 }^M A_{\m_3 , N \b} A_{\m_4 ] , P\a}
  \Theta^{Q\b} \e_{MQR} \e^{RNP}\nonumber \\
  & & \quad - \frac{1}{8} A_{[\m_1 , N\g} A_{\m_2 , P \d} A_{\m_3 ,
  Q\b} A_{\m_4 ] , M\a} \e^{\b\g} \e^{MPQ} \Theta^{N\d} ] \quad .
  \end{eqnarray}
These field strengths transform covariantly under the gauge
transformations
  \begin{eqnarray}
  & & \delta A_{\m, M\alpha } = a_{\m, M\alpha} + a_M{}^N
  A_{\m,N\alpha} -{1 \over 3} a_N{}^N A_{\m , M\alpha} + a_i
  D^i_\alpha{}^\beta A_{\m , M \beta} + \Lambda_{N\b} \Theta^{N\b} A_{\m , M\a}\nonumber \\
  & & \delta A_{\m_1 \m_2}^M = a_{\m_1 \m_2}^M -{1 \over 2}
  \epsilon^{\alpha \beta} \epsilon^{MNP} A_{[\m_1 , N\alpha } a_{\m_2]
  , P\beta} - a_N{}^M A_{\m_1 \m_2}^N + {1 \over 3 } a_N{}^N A_{\m_1 \m_2}^M + 2 \Lambda_{N\a} \Theta^{N\a} A_{\m_1 \m_2}^M\nonumber \\
  & & \delta A_{\m_1 \m_2 \m_3 , \alpha} = \de_{[\m_1} \Lambda_{\m_2 \m_3 ],\alpha} +
  A_{[\m_1 \m_2}^M a_{\m_3 ], M\alpha} -{1 \over 3!} \epsilon^{MNP }
  \epsilon^{\beta\gamma} A_{[\m_1 , M\alpha} A_{\m_2 N \beta} a_{\m_3 ]
  ,P\gamma} \nonumber \\
  & & \quad + a_i D^i_\alpha{}^\beta A_{\m_1 \m_2 \m_3 ,
  \beta}+ 3 \Lambda_{M\b} \Theta^{M\b} A_{\m_1 \m_2 \m_3 , \a} \quad , \label{gaugetransfsind=8}
  \end{eqnarray}
where the parameters $a$ are given in terms of the gauge parameters
$\Lambda$ as
   \begin{eqnarray}
   & & a_M{}^N = 3  \Lambda_{M\alpha} \Theta^{N\alpha} \nonumber
   \\
   & & a^i = -12 \Theta^{M\beta} D^i_\alpha{}^\beta
   \Lambda_{M \alpha} \nonumber \\
   & & a_{\m, M\alpha } = \partial_\m \Lambda_{M\alpha} -3
   \epsilon_{MNP } \epsilon_{\alpha\beta} \Lambda_\m^N
   \Theta^{P\beta} \nonumber \\
   & & a_{\m_1 \m_2}^M = \partial_{[\m_1} \Lambda_{\m_2]}^M  \quad .
   \end{eqnarray}
Using the formulae given in this paper and in \cite{hierarchyE11},
the reader can easily determine the field strengths and gauge
transformations for the remaining fields.

\section{D=9}
The internal symmetry of the maximal massless nine-dimensional
supergravity theory  is $\mathbb{R}^+ \times SL(2,\mathbb{R})$. The
bosonic sector of the theory contains the metric, three scalars, a
doublet and a singlet of vectors, a doublet of 2-forms and a 3-form.
The decomposition of $E_{11}$ appropriate to the nine-dimensional
theory corresponds to the deletion of nodes 9 and 11 in the Dynkin
diagram of fig. 1. The level zero generators and the positive level
form generators of rank at most four are
  \begin{equation}
  K^\m{}_\n \quad  R \quad R^i \quad R^{\m} \quad R^{\m , \alpha} \quad R^{\m_1 \m_2, \alpha}
  \quad R^{\m_1 \m_2 \m_3 }
  \quad  R^{\m_1 \m_2 \m_3 \m_4}
  \quad , \label{listofgeneratorsninedim}
  \end{equation}
where as in the previous section $\a$ denotes the $SL(2,
\mathbb{R})$ doublet and $i$ the $SL(2, \mathbb{R})$ triplet.

The non-trivial commutators involving the scalars and the positive
level generators in   eq. (\ref{listofgeneratorsninedim}) are
\cite{hierarchyE11}
  \begin{eqnarray}
  & & [ R, R^\m ] = - R^\m \nonumber \\
  & &  [ R , R^{\m , \alpha} ] = R^{\m
  ,\alpha} \nonumber \\
  & & [ R^i , R^{\m , \alpha } ] = D^i_\beta{}^\alpha R^{\m , \beta}
\nonumber \\
& &
  [ R^i , R^{\m_1 \m_2 , \alpha} ] = D^i_\beta{}^\alpha R^{\m_1 \m_2 ,
  \beta} \nonumber \\
  & &  [ R^{\m_1} , R^{\m_2 , \alpha} ] = - R^{\m_1 \m_2
  , \alpha} \nonumber \\
  & &
  [ R , R^{\m_1 \m_2 \m_3 } ] = R^{\m_1 \m_2 \m_3} \nonumber \\
  & &
  [ R^{\m_1 , \alpha} , R^{\m_2 \m_3, \beta }]= \epsilon^{\alpha \beta} R^{\m_1 \m_2 \m_3 }
\nonumber \\
  & & [ R^{\m_1 \m_2, \alpha} , R^{\m_3 \m_4 , \beta} ]= \epsilon^{\alpha \beta} R^{\m_1 \m_2
  \m_3 \m_4}
  \nonumber \\
  & & [ R^{\m_1} , R^{\m_2 \m_3 \m_4 } ]  = -R^{\m_1 \m_2 \m_3  \m_4 } \quad .
  \end{eqnarray}

We now consider the deformations of the $E_{11,9}^{local}$ algebra
as they result from applying to this case the generic commutators of
eq. (\ref{defalgebra1}) and (\ref{defalgebra2}). Imposing the
closure of the Jacobi identities involving two positive level
generators and momentum one obtains
  \begin{eqnarray}
  & & [ R^\m , P_\n ] = \delta^\m_\n [ \Theta K + \Theta R ]
  \nonumber \\
  & & [ R^{\m , \a} , P_\n ] = \delta^\m_\n [ \Theta^\a K - 4
  \Theta^\b D_{i , \b}{}^\a R^i ] \nonumber \\
  & & [ R^{\m_1 \m_2 , \a} , P_\n ] = -2 \Theta \delta^{[\m_1}_\n
  R^{\m_2 ] , \a} - \Theta^\a \delta^{[\m_1 }_\n R^{\m_2 ]}
  \nonumber\\
  & & [ R^{\m_1 \m_2 \m_3 } , P_\n ] = 3 \e_{\a \b} \Theta^\a
  \delta^{[\m_1 }_\n R^{\m_2 \m_3 ] , \b} \nonumber \\
  & & [ R^{\m_1 \m_2 \m_3 \m_4 } , P_\n ] = -4 \Theta \delta^{[\m_1
  }_\n R^{\m_2 \m_3 \m_4 ]} \quad ,
  \end{eqnarray}
where we are considering the two embedding tensors $\Theta$ and
$\Theta^\a$. The only quadratic constraint that arises from imposing
the closure of all the other Jacobi identities is
  \begin{equation}
  \Theta \Theta^\a = 0 \quad . \label{quadraticinnine}
  \end{equation}
This constraint simply means that the two embedding tensors can not
be included at the same time. It is important that the constraint
$\Theta^\a \Theta^\b D_{i ,\a\b} =0 $ does not arise. Indeed, the
symmetric product of two $SL(2, \mathbb{R})$ doublets is the
triplet, and therefore this constraint would imply that the doublet
$\Theta^\a$ should vanish identically.

In this nine-dimensional case it is also rather simple to check that
these deformations are not compatible with the deformations of ref.
\cite{hierarchyE11} corresponding to the standard gaugings of the
maximal nine-dimensional supergravity. This is exactly what happens
also in the IIA case of \cite{trombone1}, in which the trombone
deformation associated to the gauged IIA theory of
\cite{hlw,fibrebundles} is not compatible with the Romans
deformation.

We finally determine the field strengths and gauge transformations
of the fields. Starting from the group element of eq.
(\ref{generalgroupelement}) and using eqs.
(\ref{generalformoffieldstrengths}) and
(\ref{generalformofgaugetransfs}) we determine the field strength of
the vector, the 2-form and the 3-form. We consider for compactness
of notation both embedding tensors in the same equations, although
we know that because of eq. (\ref{quadraticinnine}) they can not be
simultaneously non-vanishing. The field strengths of the 1-forms are
  \begin{eqnarray}
  & & F_{\m_1 \m_2 } = 2 [ \partial_{[\m_1} A_{\m_2]} + A_{\m_1 \m_2 , \a}
  \Theta^\a - A_{[\m_1 ,\a} A_{\m_2 ]} \Theta^\a] \nonumber \\
  & & F_{\m_1 \m_2 , \alpha} = 2 [ \partial_{[\m_1} A_{\m_2] , \alpha}  +2 A_{\m_1
  \m_2, \a}\Theta - \frac{1}{2} A_{[\m_1 ,\b}A_{\m_2 ] ,
  \a}\Theta^\b -\frac{1}{2} \e^{\b\g}A_{[\m_1 ,\b}A_{\m_2 ] ,\g}
  \Theta_\a \nonumber \\
  & & \quad + \frac{1}{2} A_{[\m_1 ,\a} A_{\m_2 ] , \b} \Theta^\b ]
  \quad ,
  \end{eqnarray}
the field-strength of the 2-form is
  \begin{eqnarray}
  & & F_{\m_1 \m_2 \m_3 ,\alpha} = 3[ \partial_{[\m_1 } A_{\m_2 \m_3 ] ,
  \alpha} - A_{[\m_1} \partial_{\m_2} A_{\m_3 ], \alpha}  + 3 \e_{\a\b}\Theta^\b A_{\m_1 \m_2 \m_3}
  - A_{[\m_1 \m_2 , \b} A_{\m_3 ] ,\a}\Theta^\b \nonumber \\
  & & \quad-2 A_{[\m_1 \m_2 ,\a} A_{\m_3 ]} \Theta+
  A_{[\m_1 , \b} A_{\m_2 ,\a} A_{\m_3 ]} \Theta^\b + \frac{1}{2}
  \e^{ \b\g} A_{[\m_1 , \b }A_{\m_2 , \g} A_{\m_3 ]} \Theta_\a ]
  \quad ,
  \end{eqnarray}
and the field strength of the 3-form is
  \begin{eqnarray}
  & & F_{\m_1 ...\m_4} = 4 [ \partial_{[\m_1} A_{\m_2 ...\m_4 ]} +
  \epsilon^{\alpha\beta} A_{[\m_1 ,\a} \partial_{\m_2} A_{\m_3 \m_4 ] ,
  \beta}  + 4 \Theta A_{\m_1 ...\m_4} + 3 \Theta^\a A_{[\m_1 \m_2 \m_3}
  A_{\m_4 ] , \a} \nonumber \\
  & & \quad - \frac{1}{2} A_{[\m_1 \m_2, \a} A_{\m_3 ,\b}A_{\m_4 ]
  \g} \Theta^\a \e^{\b\g}
  ]
  \quad .
  \end{eqnarray}
These field strengths transform covariantly under the gauge
transformations
  \begin{eqnarray}
  & & \delta A_\m = a_\m  + \Lambda_\a \Theta^\a A_{\m} \nonumber \\
  & & \delta A_{\m , \alpha} = a_{\m ,\alpha}  + 2 \Lambda\Theta A_{\m ,\a} + \Lambda_\b \Theta^\b A_{\m ,\a} + a_i
  D^i_\alpha{}^\beta A_{\m ,\beta} \nonumber \\
  & & \delta A_{\m_1 \m_2 ,\alpha } = a_{\m_1 \m_2 , \alpha} + A_{[\m_1
  ,\alpha} a_{\m_2 ]} + 2 \Lambda\Theta A_{\m_1 \m_2  ,\a} + 2\Lambda_\b \Theta^\b A_{\m_1 \m_2
   ,\a}+ a_i D^i_\alpha{}^\beta A_{\m_1 \m_2 ,\beta}
  \nonumber \\
  & & \delta A_{\m_1 \m_2 \m_3 } = \de_{[\m_1} \Lambda_{\m_2 \m_3 ]} - \epsilon^{\alpha
  \beta} A_{[\m_1 \m_2 ,\alpha} a_{\m_3 ],\beta} + {1 \over 2}
  \epsilon^{\alpha \beta} A_{[\m_1 ,\alpha} A_{\m_2 ,\beta} a_{\m_3]} \nonumber \\
  & & \quad + 4 \Lambda\Theta A_{\m_1 \m_2 \m_3} + 3 \Lambda_\a \Theta^\a A_{\m_1 \m_2 \m_3} \quad ,
  \end{eqnarray}
where the parameters $a$ are expressed in terms of the gauge
parameters $\Lambda$ as
  \begin{eqnarray}
  & & a^i = - 4 \Lambda_\a \Theta^\b D^i_\b{}^\a \nonumber \\
  & & a_\m = \de_\m \Lambda - \Lambda_{\m ,\a} \Theta^\a \nonumber
  \\
  & & a_{\m , \a} = \de_\m \Lambda_{\a} -2 \Lambda_{\m ,\a} \Theta
  \nonumber \\
  & & a_{\m_1 \m_2 , \a} = \de_{[\m_1 } \Lambda_{\m_2 ] , \a} -3
  \e_{\a\b} \Theta^\b \Lambda_{\m_1 \m_2} \quad .
  \end{eqnarray}

The maximal supergravity theory in nine dimensions corresponding to
the gauging of the trombone symmetry was derived in
\cite{allgaugingsD=9}. This nine-dimensional case concludes the
analysis carried out in this paper.

\section{Conclusions}
In \cite{trombone1} a new type of  deformation of the local $E_{11}$
algebra corresponding to the IIA theory was derived, and it was
shown that it describes the gauged IIA theory of
\cite{hlw,fibrebundles}. In this paper we have generalised these
results to all dimensions, determining all possible trombone
deformations of the local $E_{11}$ algebra in $D$ dimensions. With
respect to the deformations analysed in \cite{hierarchyE11}, these
deformations not only involve the generators of the internal
symmetry, but also the trace of the $GL(D, \mathbb{R})$ generators.
The deformations are parametrised by a constant quantity in the same
representation ${\bf R_1}$ of $E_{11-D}$ as the 1-form generators.
This quantity is identified with the embedding tensor introduced in
\cite{henning} to describe the gauge algebra of the maximal
supergravity theories in which the scaling symmetry is gauged in
dimension from three to six. We have determined the quadratic
constraints resulting from closure of the Jacobi identities, which
coincide with the quadratic constraints of \cite{henning}. We have
also determines the field strengths and gauge transformations for
the 1-forms, 2-forms and 3-forms of all theories.

The deformed algebra can naturally be extended to include higher
rank form generators, and we expect the field equations to arise as
duality relations between the corresponding field strengths. It is
important to observe, though, that the $D-1$-form generators that
are present in the decomposition of $E_{11}$ corresponding to the
$D$ dimensional theory are in the same representations of the
embedding tensors corresponding to the gauging of a subgroup of the
internal symmetry \cite{fabiopeterE11origin,ericembeddingtensor}.
Therefore, there is no form generator in the spectrum associated to
this trombone embedding tensor. As already discussed in section 2,
in \cite{henning} it was conjectured that the generators in the
representation $GL(D, \mathbb{R})$ of mixed symmetry that we denote
by $(1,D-2)$, that is the generators $R^{\m, \n_1 ...\n_{D-2} ,
M_1}$, that are in the same $E_{11-D}$ representation as the 1-form
generators and thus as the trombone embedding tensor, might trigger
these deformations. Our point of view, though, is that because the
theories considered in this paper do not admit a lagrangian
formulation, the fact that there are no form generators associated
to these deformations is completely consistent, and we do not expect
any $E_{11}$ generator associated to a non-propagating field to play
a role in triggering these deformations.

In two dimensions the scaling symmetry becomes an off-shell
symmetry, that is the central extension of the affine internal
symmetry group $E_{9(9)}$ \cite{juliainfinite}. As a consequence, as
stressed in \cite{henning}, the embedding tensor associated to the
trombone gauging and the one associated to the internal gauging
coincide. This is completely consistent from the $E_{11}$ point of
view. Indeed the embedding tensor associated to the trombone gauging
belongs to the ${\bf R_1}$ representation of $E_{11-D}$, that is the
representation to which the 1-form generators belong, while the
embedding tensor associated to the internal gauging belongs to the
${\bf R_{D-1}}$ representation of the $D-1$ form generators
\cite{fabiopeterE11origin,ericembeddingtensor}. In $D=2$ the 1-forms
and the $D-1$ forms coincide. The $E_{11}$ decomposition associated
to the two-dimensional theory results from deleting node 2 in the
Dynkin diagram of fig. 1. From the diagram it is manifest that the
1-form generators belong to the $E_9$ representation with $p_3 =1$,
where $p_3 $ is the Dynkin index associated to node 3 in the
diagram. The gaugings of the maximal supergravity theory in two
dimensions using the embedding tensor formalism were derived in
\cite{samtlebenweidnerD=2}, and  it would be interesting to analyse
this from the $E_{11}$ perspective.

Finally, it is worth mentioning that the algebraic construction
derived in this paper can be extended to other supergravity theories
with less supersymmetry whose bosonic sector still admits a
description in terms of a very-extended Kac-Moody algebra. In
particular, all theories with 16 supercharges are described in terms
of very-extended algebras of $B$ or $D$ type
\cite{igorpeter16super}, while in \cite{fabiopetertoine} it was
shown that all theories with eight supersymmetries whose reduction
to three dimensions gives rise to scalars that parametrise symmetric
manifolds correspond to non-linear realisations of very-extended
Kac-Moody algebras for suitable choices of real forms. The same
applies to all supergravity theories with more than 16 supercharges
\cite{fabiopetertoine}. For all these theories we expect that the
trombone gaugings are associated to deformations of the
corresponding very-extended algebras as described in section 2,
hence the title of this paper.

\vskip 2cm

\section*{Acknowledgments}
I would like to thank the organisers of the FPUK v3.0 conference in
Cambridge for creating a stimulating environment while this project
was at its early stages. This work is supported by the PPARC rolling
grant PP/C5071745/1, the EU Marie Curie research training network
grant MRTN-CT-2004-512194 and the STFC rolling grant ST/G000/395/1.

\end{document}